\documentclass[11pt, a4paper]{article}
\usepackage{jheppub}
\usepackage{bbm}
\usepackage{amsmath}
\usepackage{amssymb}
\usepackage{amsfonts}
\usepackage{mathrsfs}
\usepackage{graphicx}
\usepackage[11pt]{moresize}
\usepackage[normalem]{ulem}
\usepackage[dvipsnames]{xcolor}
\usepackage{hyphenat}
\usepackage[]{hyperref}
\usepackage{orcidlink}
\usepackage{slashed}
\usepackage{relsize}
\usepackage{adjustbox}
\usepackage[verbose]{placeins}
\usepackage{multirow}
\usepackage{float}
\usepackage{footnote}
\usepackage{tablefootnote}
\usepackage{lineno}
\usepackage{makecell}
\usepackage{subfigure}
\usepackage{comment}

\hypersetup{
 bookmarks  = true,      
 unicode    = false,     
 pdfcreator = {RevTeX},  
 colorlinks = true,      
 linkcolor  = red,       
 citecolor  = blue,      
 filecolor  = black,     
 urlcolor   = blue,      
}

\makeatletter
\renewcommand{\fnum@table}{\textbf{\tablename~\thetable}}
\renewcommand{\fnum@figure}{\textbf{\figurename~\thefigure}}

\makeatother

\newenvironment{conditions*}
{\par\vspace{\abovedisplayskip}\noindent
  \tabularx{\columnwidth}{>{$}l<{$} @{${}={}$} >{\raggedright\arraybackslash}X}}
{\endtabularx\par\vspace{\belowdisplayskip}}

\preprint{}

\title{Exploring new physics with DUNE high energy flux: the case of Lorentz
Invariance Violation, Large Extra Dimensions and Long Range Forces} 

\author[a,b]{Alessio Giarnetti,}
\author[a,b]{Simone Marciano,}
\author[a,b]{Davide Meloni}

\affiliation[a]{Dipartimento di Matematica e Fisica, Universit\'a di Roma Tre,
Via della Vasca Navale 84, 00146, Roma, Italy}

\affiliation[b]{INFN Sezione di Roma Tre, Via della Vasca Navale 84, 00146, Roma, Italy}

\abstract
{DUNE is a next-generation long-baseline neutrino oscillation experiment. It is expected to measure with an unprecedent precision the atmospheric oscillation parameters, including the CP-violating phase $\delta_{CP}$. Moreover, several studies have suggested that its unique features should allow DUNE to probe several new physics scenarios. In this work, we explore the performances of the DUNE far detector in constraining new physics if a high-energy neutrino flux is employed (HE-DUNE). We take into account three different scenarios: Lorentz Invariance Violation (LIV), Long Range Forces (LRF) and Large Extra Dimensions (LED). Our results show that HE-DUNE should be able to set bounds competitive to the current ones and, in particular, it can outperform the standard DUNE capabilities in constraining CPT-even LIV parameters and the compactification radius $R_{ED}$ of the LED model. 
}


\begin{document}
\maketitle
\flushbottom

\section{Introduction}
\label{sec:introduction}

 Neutrino oscillation discovery \cite{Super-Kamiokande:1998kpq} represented a milestone in the history of particle physics. Indeed, the observation of this phenomenon unveiled that neutrino have (tiny) masses compared to the other fermions of the Standard Model (SM) of particle physics. 
 In addition, the small uncertainties achieved in the measurements of mixing angles carry us into a precision era in the neutrino sector, thanks to an effort lasted 25 years and that has involved
 different particle sources: the Sun \cite{XU2023104043}, the Earth atmosphere \cite{CHOUBEY2016235}, nuclear reactors \cite{Cao:2017drk} and accelerator facilities \cite{Mezzetto:2020jka}. The oscillation parameters involved in solar oscillation, namely the solar mixing angle $\theta_{12}$ and the solar mass splitting $\Delta m_{21}^2$, have been precisely measured by several solar neutrino experiment and by a peculiar reactor experiment: KamLAND \cite{KamLAND:2008dgz,KamLAND:2010fvi,KamLAND:2013rgu}. The reactor mixing angle $\theta_{13}$, instead, was discovered and measured with an astonishing precision in 2012 \cite{DayaBay:2012fng}, leaving the atmospheric oscillation sector the less constrained. Indeed, the  mixing angle $\theta_{23}$ is almost maximal ($\theta_{23}\in [40^\circ,50^\circ]$ \cite{Esteban:2020cvm}) and suffers from the so-called {\it octant degeneracy} which makes the determination of the octant in which $\theta_{23}$ lies (Higher Octant, HO, $\theta_{23}>45^\circ$ or Lower Octant, LO, $\theta_{23}<45^\circ$) very difficult to be determined. Moreover, according to the current neutrino oscillation data, the atmospheric mass splitting $\Delta m_{31}^2$ has an absolute value 30 times larger than the solar one and can still assume both positive and negative values. This is the so-called {\it mass hierarchy problem}. \\
The solution of these long-lasting problems might be given by the next-generation of long baseline (LBL) accelerator experiments. Such experiments are mainly sensitive to the atmospheric neutrino frequency and  employ a well-known, focused, artificial muon neutrino flux coming from an accelerator facilities. The two experiments which are expected to start their data taking in the current decade are T2HK in Japan \cite{Hyper-Kamiokande:2018ofw} and DUNE \cite{DUNE:2015lol} in the USA. The importance of such experiments lies not only on the solution of the octant and hierarchy problems, but also on their unprecedent capabilities to measure the CP-violating phase $\delta_{\rm CP}$, for which T2K and NO$\nu$A \cite{Himmel:2020kct,Batkiewicz-Kwasniak:2022vrj,Rahaman:2021zzm} have provided a first (weak) signal. Even though both experiments will be sensitive to the same neutrino oscillation regime, DUNE will have the advantage to run with a broad-band beam, allowing to observe neutrinos whose energy extends beyond the first  oscillation maximum. 
To go even higher in energy, the possibility to use an High-Energy (HE) neutrino beam in DUNE has been widely discussed. This would allow, for instance, to collect the largest $\nu_\tau$ events sample ever observed \cite{Ghoshal:2019pab}. In the last years, a large number of studies have shown that the employment of this particular flux might be extremely useful in exploring tiny new physics effects in neutrino oscillation \cite{Masud:2017bcf, Ghoshal:2019pab, DeGouvea:2019kea,DeRomeri:2023dht,MammenAbraham:2022xoc}. 
In this context, the advantage of using the HE flux  lies on the fact that some models predicts  new  effects enhanced by neutrino energy. The extremely large collectable sample of events in all neutrino flavors provides an unique tool to study these energy-enhanced new physics effects. \\
In this paper we explore and discuss the DUNE capabilities in its HE configuration to measure the new physics parameters involved in neutrino oscillation in three particular models: the Lorentz Invariance Violation (LIV), the Long Range Forces (LRF) and the Large Extra Dimensions (LED) models. In the first two cases, the interaction Lagrangian is supplemented with additional operators whose coefficients must be bounded from above, while in the last case the space-time framework in enlarged with at least one spatial dimension (experienced by right-handed neutrinos only) whose radius belong to the sub-millimeter range.   
Even though such New Physics models have been investigated by other authors in the literature, our results show that the bounds HE-DUNE can set on the model parameters are comparable with the existing ones and, for the CPT-even LIV parameters and for the radius of the LED model, can outperform the capabilities of the standard DUNE set-up.

The paper will be organized as follows: in Sec.~\ref{Sec:HEDUNE} we will describe the DUNE experiment and its HE configuration, in Secs.~\ref{sec:LIVfull},~\ref{sec:LRFfull} and~\ref{sec:LEDfull} we discuss the LIV, LRF and LED models, respectively. Finally, in Sec.~\ref{sec:concl}, we draw our conclusions.

\section{The DUNE experiment and the High-Energy flux}
\label{Sec:HEDUNE}

The DUNE (Deep Underground Neutrino Experiment) experiment is a proposed long-baseline experiment in the USA \cite{DUNE:2015lol,DUNE:2016hlj,DUNE:2020jqi,DUNE:2020ypp}. The near detector complex, composed by different multi-purpose near detectors \cite{DUNE:2021tad} as well as the accelerator facilities, are being built at Fermilab; on the other hand, the 40kt LAr-TPC detector will be located in South Dakota, roughly 1300 km far from the neutrino beam source. The on-axis neutrino beams will be mainly composed by $\nu_\mu$ or $\bar\nu_\mu$ depending on the current circulating in the focusing horns; this will allow the experiment to run in both \textit{neutrino} and \textit{antineutrino} modes. The main purpose of the experiment is to precisely measure the oscillation parameters in the atmospheric sector. In particular, DUNE is expected to reach a great sensitivity to the mass hierarchy and an unprecedent sensitivity to the $\theta_{23}$ octant. Moreover, the experiment is also designed to maximize the sensitivity to the PMNS matrix phase $\delta_{CP}$. In order to perform such a measurement, the proposed neutrino flux is a broad-band beam peaked at around 2.5 GeV so to sit at the first atmospheric oscillation maximum. This should allow to observe not only a huge sample of $\nu_\mu$ events ($\mathcal{O}(10^3)$ per year), but also several $\nu_e$ appearance events ($\mathcal{O}(10^2)$ per year) \cite{DUNE:2015lol,DUNE:2016hlj,DUNE:2020jqi,DUNE:2020ypp}. A very intense flux of $\nu_\tau$ will also arrive at the far detector; however, given the Charged Current (CC) $\nu_\tau$ interaction energy threshold (3.1 GeV), only a minor fraction of these events might be observed. To overcome this problem and observe a larger number of $\tau$ neutrinos a broader, high energetic flux peaked at around 5 GeV has also been considered by the DUNE collaboration \cite{HEDUNE-1,HEDUNE-2,HEDUNE-3}. The main disadvantage of using this flux is that the performances in measuring standard oscillation parameters are poorer \cite{Ghoshal:2019pab}. However, other than collecting the larger $\nu_\tau$ sample ever obtained ($\mathcal{O}(10^2)$ per year \cite{Ghoshal:2019pab}), the employement of such a flux has been demonstrated to be very promising in constraining new physics scenarios \cite{Masud:2017bcf,Ghoshal:2019pab,DeGouvea:2019kea,DeRomeri:2023dht}. In this work, we focus on the capabilities of the DUNE experiment to probe some Beyond the Standard Model (BSM) theories which make an imprint on the neutrino oscillation probabilities, taking full advantage of the high energy flux. From now on, we will refer to this DUNE configuration as HE-DUNE. In order to make a comparison with the standard DUNE results, for the HE-DUNE  we will use  the same efficiencies, energy resolutions and systematic uncertainties provided by the collaboration for the standard DUNE \cite{DUNE:2016ymp,DUNE:2021cuw}. In order to include the possible effects of the $\nu_\tau$ appearance in constraining the new physics parameters space, we follow \cite{Ghoshal:2019pab} and \cite{DeGouvea:2019kea}. In particular, we make the hypothesis that the HE-DUNE might be able to observe 30\% of the $\nu_\tau$ events with subsequent $\tau\to e$ decays and 30\% of the $\nu_\tau$ events with subsequent hadronic $\tau$ decays. The systematic uncertainty for this channel has been set to a conservative 25\% normalization error\footnote{For the other two oscillation channels, namely $\nu_e$ appearance and $\nu_\mu$ disappearance, the collaboration suggested 2\% and 5\%, respectively.}. Misidentified $\nu_\mu$, $\nu_e$ and NC events have been considered as a background to the $\nu_\tau$ channel, according to \cite{Ghoshal:2019pab,DeGouvea:2019kea}. \\
The total running time for HE-DUNE has been fixed to 5 years in neutrino mode and 5 years in antineutrino mode. Finally, the whole analysis have been carried out neglecting the events observed at the Near Detectors and considering the final 40kt far detector for all the 10 years of data taking.

\section{The Lorentz Invariance Violation case}
\label{sec:LIVfull}

The Lorentz invariance is one of the fundamental symmetries of the SM and it is related to the space-time structure. The other essential symmetry of the quantum field theory is the CPT symmetry\footnote{In Quantum Field Theory the \emph{CPT theorem} states that the combination of the discrete transformations “Charge\hyp{}conjugation” (C), “Parity” (P) and “Time\hyp{}reversal” (T) must be a symmetry
of the theory.}. Since the SM fails to unify all the forces governing the Universe, it has been taken into account the possibility that the SM is an effective theory of a wider framework which unifies not only electromagnetic and weak interactions, but also strong interactions and gravity. The energy scale of such a general theory should be the Planck mass ($M_P\sim10^{19}$ GeV). In these SM extensions, CPT and Lorentz symmetries might be violated \cite{Colladay:1996iz,Colladay:1998fq,Diaz:2011ia,Greenberg:2002uu,Kostelecky:1991ak,Kostelecky:1994rn,Kostelecky:1995qk,Kostelecky:2003cr,PhysRevD.39.683,PhysRevLett.63.224}; in particular, it has been shown that CPT breaking always leads to Lorentz Invariance Violation (LIV) as well \cite{Greenberg:2002uu}. Neutrino experiments could be able to probe LIV through suitable  modification of the oscillation probabilities  induced by the presence of new terms in the full theory. In the next subsection we will show how the neutrino probabilities can be affected by LIV.  

\subsection{Theoretical Framework}

In the presence of LIV, the neutrino lagrangian density term can be written as \cite{Kostelecky:2003cr,Agarwalla:2023wft,KumarAgarwalla:2019gdj}
\begin{equation}
    \mathcal{L}=\frac{1}{2}\bar{\psi}(i\slashed{\partial}-M+Q)\psi+h.c\,,
\end{equation}
where $\psi$ is the neutrino fermionic field and the effect of the LIV is encoded in the generic operator Q. 
If we restrict ourselves to renormalizable Dirac  couplings, the Lorentz Invariance violating lagrangian terms can be written as \cite{Kostelecky:2011gq,KumarAgarwalla:2019gdj}:
\begin{equation}
    \mathcal{L}^{\text{LIV}}=-\frac{1}{2}\left(a^{\mu}_{\alpha\beta}\bar\psi_\alpha \gamma_\mu \psi_\beta+b^{\mu}_{\alpha\beta}\bar\psi_\alpha \gamma_5\gamma_\mu \psi_\beta-ic^{\mu\nu}_{\alpha\beta}\bar\psi_\alpha \gamma_\mu\partial_\nu \psi_\beta -id^{\mu\nu}_{\alpha\beta}\bar\psi_\alpha \gamma_5\gamma_\mu\partial_\nu \psi_\beta \right)+h.c.\,.
\end{equation}
The first two terms are CPT-odd, while the third and the fourth terms are CPT-even. Thus, the LIV effect  in the interaction hamiltonian can be encoded in two hermitian matrices:
\begin{equation}
    \tilde {a}^\mu_{\alpha\beta}=(a+b)^\mu_{\alpha\beta} \quad \quad \text{and} \quad \quad \tilde{c}^{\mu\nu}_{\alpha\beta}=(c+d)^{\mu\nu}_{\alpha\beta} \,.
\end{equation}
These matrices modify the standard neutrino oscillation Hamiltionian, by adding a new term $H_{\rm{LIV}}$ to the vacuum and matter contributions:
\begin{equation}
    H_{\nu}=H_{\rm{m}}+H_{\rm{\rho}}+H_{\rm{LIV}} \,,
\end{equation}
where, as usual, 
\begin{equation}
H_{\rm{m}}=\frac{1}{2E} U\begin{pmatrix}
    0 & 0 & 0 \\
    0 & \Delta m_{21}^2 & 0 \\
    0 & 0 & \Delta m_{31}^2
\end{pmatrix} U^\dagger \quad \text{and}  \quad
H_{\rm{\rho}}=\sqrt{2}G_F N_e\text{diag}(1,0,0)\,.
\label{eq:stdnuH}
\end{equation}
Here, $E$ is the neutrino energy, $\Delta m_{ij}^2=m^2_{i}-m^2_j$ are the neutrino mass splittings, $U$ is the neutrino mixing matrix which depends on three mixing angles ($\theta_{12}$, $\theta_{13}$ and $\theta_{23}$) and one complex phase $\delta_{\rm{CP}}$, $G_F$ is the Fermi constant and $N_e$ is the number density of electrons in the medium traversed by neutrinos. The term $H_{\rm{\rho}}$ corresponds to the well-known Mikheev-Smirnov-Wolfenstein (MSW) mechanism \cite{PhysRevD.17.2369},

The last term $H_{\rm{LIV}}$ is the one induced by the LIV. On a general ground, it reads $H_{\rm{LIV}}=1/E (\tilde{a}^\mu p_\mu-\tilde{c}^{\mu\nu}p_\mu p_\nu)$, where $p_\mu$ is the neutrino four-momentum. However, focusing only on time-like LIV matrices  components ($\mu,\nu=0$) and considering a Sun-centered isotropic inertial frame (see \cite{Kostelecky:2011gq} for details), the Lorentz Invariance Violation effects are governed by the parameters $\tilde{a}^{0}_{\alpha\beta}\equiv a_{\alpha\beta}$ and $\tilde{c}^{00}_{\alpha\beta}\equiv c_{\alpha\beta}$. Being elements of hermitian matrices, diagonal $(a,c)_{\alpha\alpha}$  are real while off-diagonal $(a,c)_{\alpha\beta}$ are complex parameters uniquely determined by their moduli, which we denote as $(a,c)_{\alpha\beta}$, and their phases $\Phi_{\alpha\beta}$. Thus, the LIV Hamiltonian reads:
\begin{equation}
    H_{\rm{LIV}}=\begin{pmatrix}
        a_{ee} & a_{e\mu} & a_{e\tau} \\
        a_{e\mu}^* & a_{\mu\mu} & a_{\mu\tau} \\
        a_{e\tau}^* & a_{\mu\tau}^* & a_{\tau\tau}
    \end{pmatrix}-\frac{4}{3} E \begin{pmatrix}
        c_{ee} & c_{e\mu} & c_{e\tau} \\
        c_{e\mu}^* & c_{\mu\mu} & c_{\mu\tau} \\
        c_{e\tau}^* & c_{\mu\tau}^* & c_{\tau\tau}\,,
    \end{pmatrix}
\end{equation}
where, as already mentioned, $a_{\alpha\beta}$ ($c_{\alpha\beta}$) are CPT-odd (CPT-even) LIV parameters. Notice that the factor $-4/3$ comes from the fact that the trace of $\tilde{c}$ is not observable and its diagonal space components must be related to its 00 component \cite{Kostelecky:2011gq}. It is worth to mention that $a_{\alpha\beta}$ matrix has a similar structure to  the propagation Non Standard Interaction (NSI) matrix.
Even though there exists a direct correspondence between NSI and LIV parameters, it is important to notice that the two models affect neutrino oscillation in a different way: while NSIs necessitate neutrinos to travel through a matter medium, LIV modifies the oscillation probabilities also in vacuum \cite{Agarwalla:2023wft,KumarAgarwalla:2019gdj,Barenboim:2018lpo,Diaz:2015dxa}. \\
Another important aspect of LIV effects is that the CPT-even ones are amplified by neutrino energy. For this reason, we expect an high-energy flux for the DUNE experiment to be more efficient in constraining them. From now on, we will only focus on the off-diagonal LIV parameters since they affect the most the oscillations of neutrinos at a long-baseline experiment, as we will discuss below. \\ 
In order to have a feeling of the effect of the LIV on the oscillation parameters, we can write the correction to the $\nu_\mu\to\nu_e$ and $\nu_\mu\to\nu_\mu$ SM  probabilities at the first order in $a_{\alpha\beta}$ and $c_{\alpha\beta}$ as \cite{Agarwalla:2023wft}
\begin{equation}
\begin{aligned}
    P_{\mu e}^{\rm{LIV}}\sim & 2 L s_{13}\sin 2\theta_{23}\sin\Delta\bigg\{\mathcal{F}_{e\mu} \bigg[-c_{23} \sin\Delta\sin(\delta_{CP}+\Phi_{e\mu})+\\
    +&c_{23}\left(\frac{s_{23}^2\sin\Delta}{c_{23}^2 \Delta}+\cos\Delta\right)\cos(\delta_{CP}+\Phi_{e\mu}) \bigg]+\\
   +&\mathcal{F}_{e\tau} \left[s_{23} \sin\Delta\sin(\delta_{CP}+\Phi_{e\tau})+s_{23}\left(\frac{\sin\Delta}{\Delta}-\cos\Delta\right)\cos(\delta_{CP}+\Phi_{e\tau}) \right] \bigg\}  \,,
\end{aligned}
\label{eq:PMUE}
\end{equation}
where $s_{ij}=\sin\theta_{ij}$, $c_{ij}=\cos\theta_{ij}$, $\Delta=\Delta m_{31}^2 L/4E$ with L being the distance travelled by the neutrino and finally $\mathcal{F}_{\alpha\beta}$ is either $|a_{\alpha\beta}|$ or $-4/3 E |c_{\alpha\beta}|$ (the related phases are indicated with $\Phi_{\alpha\beta}$). It is clear that the leading corrections to the standard $P_{\mu e}$  are driven by the $e-\mu$ and $e-\tau$ LIV parameters. For the muon neutrino disappearance, instead, we obtain \cite{Agarwalla:2023wft}:
\begin{equation}
\begin{split}
    P_{\mu \mu}^{\rm{LIV}}\sim & \frac{1}{2}\sin^2 2\theta_{23}\left[2\Delta\sin^2\theta_{13}-2L\sin 2\theta_{23}\mathcal{F}_{\mu\tau} \cos\Phi_{\mu\tau}\right] \,;
\end{split}
\label{eq:MUMU}
\end{equation}
the main dependence on the LIV parameters is given by $a_{\mu\tau}$ ($c_{\mu\tau}$)\footnote{The $\nu_\tau$ appearance probability can be obtained from unitarity and the leading term will be the one depending again on $a_{\mu\tau}$ ($c_{\mu\tau}$).}. \\
Several bounds on LIV parameter have been obtained using long baseline accelerator neutrinos \cite{Barenboim:2018ctx,Agarwalla:2023wft,KumarAgarwalla:2019gdj,Fiza:2022xfw,MINOS:2008fnv,MINOS:2010kat,MINOS:2012ozn,Dighe:2008bu,Barenboim:2009ts,Rebel:2013vc,Diaz:2015dxa,deGouvea:2017yvn,Barenboim:2017ewj,Majhi:2019tfi,Majhi:2022fed}, short baseline accelerator neutrinos \cite{T2K:2017ega,MiniBooNE:2011pix}, reactor neutrinos \cite{Giunti:2010zs,DoubleChooz:2012eiq}, solar neutrinos \cite{Diaz:2016fqd}, high energy astrophysical neutrinos \cite{Hooper:2005jp,Tomar:2015fha,Liao:2017yuy,IceCube:2010fyu} and  atmospheric neutrinos \cite{Chatterjee:2014oda,Sahoo:2021dit,Datta:2003dg,Super-Kamiokande:2014exs} (see also  \cite{Arguelles:2019xgp,Arguelles:2022tki} for reviews on this, and other, new physics models). Since  we are interested in the DUNE experiment performances with an high-energy flux, in the following we summarize the bounds that DUNE in its standard configuration \cite{DUNE:2020jqi,DUNE:2020ypp,DUNE:2021cuw} might set on off-diagonal LIV parameters. In \cite{Agarwalla:2023wft} the authors found, at 95\% CL:
\begin{equation}
    \begin{aligned}
        |a_{e\mu}|&< 1.00 \times 10^{-23} \quad \rm{GeV}\,,\quad  |c_{e\mu}|< 0.66 \times 10^{-24}\,, \\
        |a_{e\tau}|&< 1.05 \times 10^{-23} \quad \rm{GeV}\,,\quad |c_{e\tau}|<  1.65 \times 10^{-24}\,, \\
        |a_{\mu\tau}|&< 1.26 \times 10^{-23} \quad \rm{GeV}\,,\quad  |c_{\mu\tau}|< 0.97 \times 10^{-24}\,,
    \end{aligned}
\end{equation}
\noindent
which are one order (two orders) of magnitude stronger than the bounds set by the current LBL experiments NO$\nu$A and T2K on CPT-violating (CPT-conserving) LIV parameters\footnote{Notice that limits from the more energetic atmospheric and astrophysical neutrinos on CPT-even LIV parameters are more stringent than the DUNE ones, due to the dependence on the neutrino energy of their effect on the oscillation probabilities.}. For $a_{e\mu}$ and $a_{e\tau}$, the authors of \cite{Agarwalla:2023wft} observed that there exists a strong correlation among them and the standard $\theta_{23}$ and $\delta_{CP}$. This allows for a second minimum in the sensitivity analysis; however, given that we fix the atmospheric mixing angle to the lower octant, we will not be able to observe such a behaviour. Neglecting the degeneracies, the limits obtained in \cite{Agarwalla:2023wft} for the standard DUNE for $a_{e\mu}$ and $a_{e\tau}$ are as follows:
\begin{equation}
    \begin{aligned}
        |a_{e\mu}|&< 3.0 \times 10^{-24} \quad \rm{GeV}\,, \\
        |a_{e\tau}|&< 4.5 \times 10^{-24} \quad \rm{GeV}\,.
    \end{aligned} 
\end{equation}

\subsection{HE-DUNE results}
\label{sec:LIV}

\begin{figure}[H]
    \centering
    \includegraphics[width=0.99\textwidth]{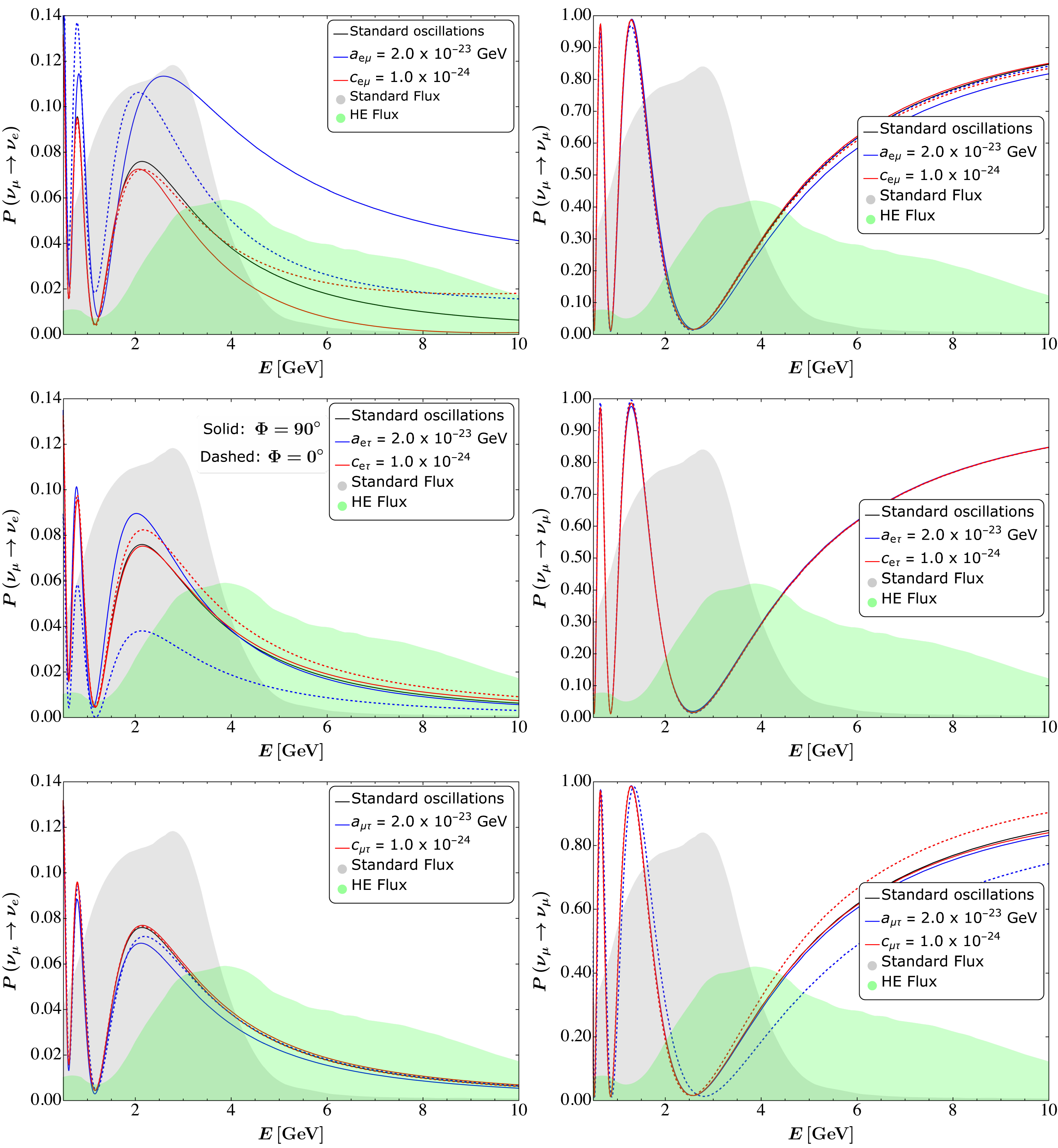}
    \caption{$\nu_e$ appearance (left panels) and $\nu_\mu$ disappearance (right panels) probabilities in presence of off-diagonal CPT violating and conserving LIV parameters. In particular top, middle and bottom panels show the effect of $a_{e\mu}$ ($c_{e\mu}$), $a_{e\tau}$ ($c_{e\tau}$) and $a_{\mu\tau}$ ($c_{\mu\tau}$), respectively.  Black lines correspond to the standard oscillation case, blue (red) lines to the probabilities obtained for $a_{\alpha\beta}=2\times10^{-23}$ GeV ($c_{\alpha\beta}=1.0\times10^{-24}$). Solid and dashed curves depict the effects of LIV phases (generically indicated $\Phi$) when $\Phi=90^\circ$ and $\Phi=0^\circ$, respectively. The grey and green shadowed regions illustrate the standard and the high-energy DUNE flux.
    }
    \label{fig:probLIV}
\end{figure}

The HE configuration of the DUNE experiment, as already discussed in Sec.~\ref{Sec:HEDUNE}, might allow accelerator neutrino energies from roughly 1 to 15 GeV. In Fig.~\ref{fig:probLIV} we show the electron appearance (left) and muon disappearance (right) oscillation probabilities with and without LIV. Black lines depict the SM oscillation probabilities computed using the best fits summarized in Tab.~\ref{tab:t1} \cite{Esteban:2020cvm,nufit}. Blue lines have been obtained setting CP-odd LIV parameters $a_{e\mu}$, $a_{e\tau}$, $a_{\mu\tau}$ to $2.0\times10^{-23}\rm{ GeV}$ in top, middle and bottom panels, respectively. Red lines show the effect of $c_{e\mu}$, $c_{e\tau}$, $c_{\mu\tau}$ with a magnitude of $10^{-24}$. We also considered two extreme values of the corresponding LIV phase $\Phi_{\alpha\beta}$, namely $0^\circ$ (dashed lines) and $90^\circ$ (solid lines). The benchmark values of the LIV parameters have been chosen of the same order of magnitude of the DUNE limits obtained with the standard neutrino flux. The shaded regions correspond to the unoscillated standard (grey) and HE (green) DUNE fluxes in arbitrary units. It is clear that in the appearance case, the most important CPT-odd LIV parameters are $a_{e\mu}$ and $a_{e\tau}$, as clearly visible in Eq.~\ref{eq:PMUE}. From the same equation, one can observe that, at the first oscillation maximum, the correction proportional to $a_{e\mu}$ is positive for both considered values of the LIV phase; on the other hand, it has a plus (minus) sign for $\Phi_{e\tau}=90^\circ$ ($\Phi_{e\tau}=0^\circ$) when the $a_{e\tau}$ parameter is taken into account. Regarding CPT-even LIV parameters, their effects on $P_{\mu e}$ become more important for higher neutrino energies, above $\sim 4$ GeV. Such an energy is located at the high-energy tail of the standard flux and at the peak of the HE flux. For this reason, we expect HE-DUNE to be more sensitive to energy-enhanced CPT-even LIV parameter than DUNE in its standard configuration. We also observe that the most relevant parameter in this case is $c_{e\mu}$. \\
For the disappearance probability, as already mentioned and explicitly shown in Eq.~\ref{eq:MUMU}, the most important LIV parameters are $a_{\mu\tau}$ and $c_{\mu\tau}$. From Fig. \ref{fig:probLIV} we can also observe a slight sensitivity of this channel to the $e-\mu$ parameters, especially in the high-energy region mostly covered by the HE-DUNE.
{\color{red}\renewcommand{\arraystretch}{1.5}
\begin{table}
\centering
\begin{tabular}{|c |c| c|} 
 \hline
Oscillation parameters ($3\nu$) & Normal ordering (NO) \\ [0.5ex] 
 \hline\hline
$\theta_{12}^{\circ}$ & $33.41^{+0.75}_{-0.72}$\\
 \hline
$\theta_{23}^{\circ}$ & $42.2^{+1.1}_{-0.9}$\\
\hline
$\theta_{13}^
{\circ}$ & $8.58^{+0.11}_{-0.11}$\\
\hline
$\delta_{CP}^{\circ}$ & $232^{+36}_{-26}$ \\
\hline
$\Delta m_{21}^2$ (eV$^2$) & $7.41^{+0.21}_{-0.20}\times 10^{-5}$ \\
\hline
$\Delta m_{31}^2$ (eV$^2$) & $+2.507^{+0.026}_{-0.027}\times 10^{-3}$ \\
\hline
\end{tabular}
\caption{ Best-fit value of the neutrino oscillation parameters in the standard three-flavour scenario. The values of the mixing angles and the mass splittings and their $1\sigma$ uncertainty intervals are taken from Ref.~\cite{Esteban:2020cvm}.}    \label{tab:t1}
\end{table}
\renewcommand{\arraystretch}{1}}\noindent

We now study the performances of DUNE in its high-energy configuration in constraining the LIV parameters. In order to perform our numerical analysis, we used the GLoBES software \cite{Huber:2004ka,Huber:2007ji} and its new physics tool \cite{Kopp:2007ne}. All the results have been obtained using a Poissonian $\chi^2$ defined as:
\begin{equation}
    \chi^2(\Vec{\lambda},a)=\sum_{i=1}^n 2\left((1+a)T_i-O_i+O_i\log\frac{O_i}{(1+a)T_i}\right)+\frac{a^2}{\sigma_a^2}\,,
\end{equation}
with $\Vec{\lambda}$ is the set of input oscillation parameters, $\sigma_a$ is the normalization error, $n$ is the number of energy bins that we fixed to 80 \cite{DUNE:2021cuw}, $O_i$ are the observed rates and $T_i$ are the theoretical rates employed in the fit. The systematic uncertainties are included using the the \textit{pull-method} described in \cite{Huber:2002mx,Fogli:2002pt}.\raggedbottom

\begin{figure}[H]
    \centering
    \includegraphics[width=0.9\textwidth,height=20cm]{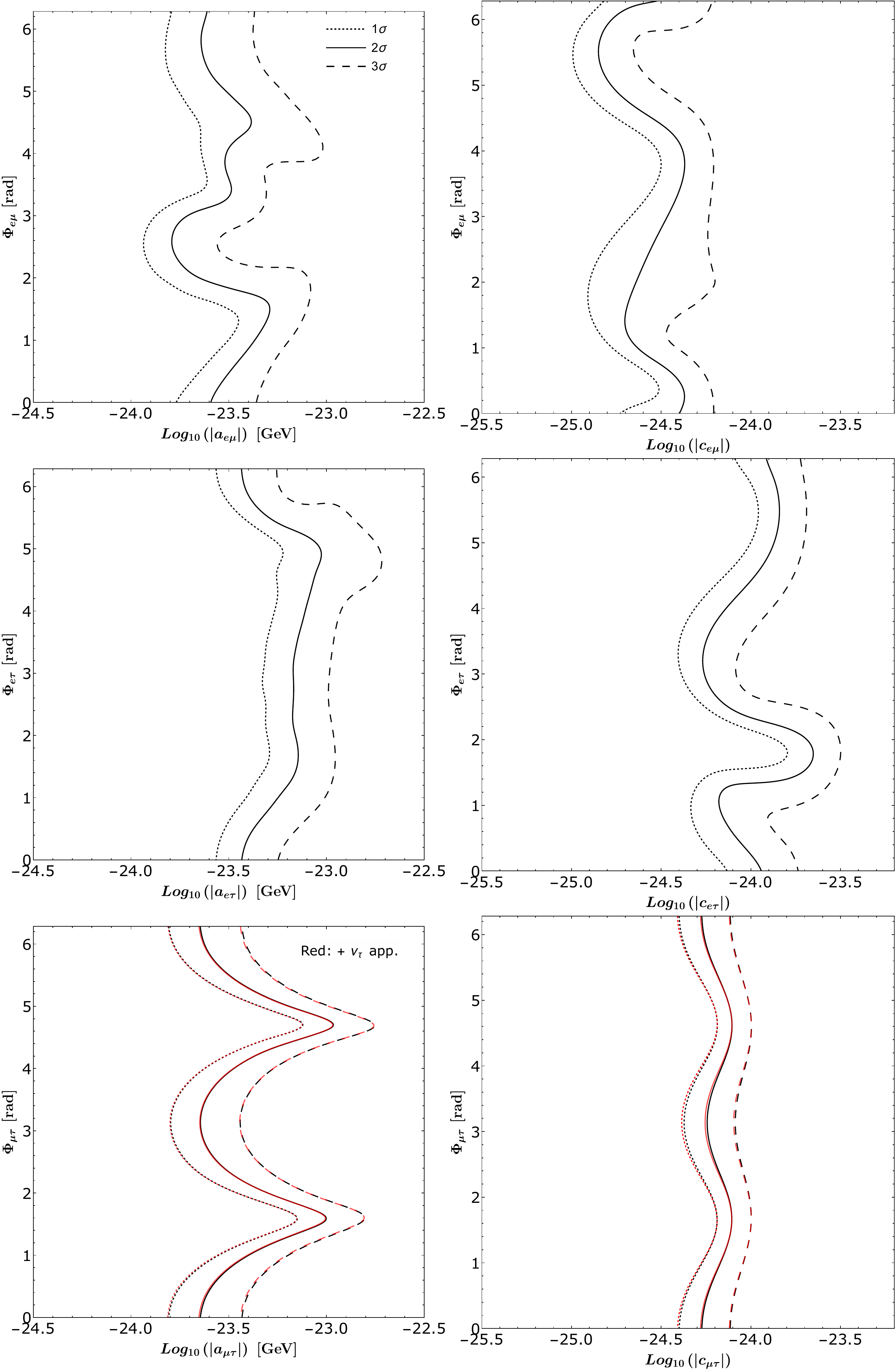}
    \caption{$1\sigma$ (dotted), $2\sigma$ (solid) and $3\sigma$ (dashed) allowed contours in the $|a_{\alpha\beta}|-\Phi_{\alpha\beta}$ (left panels) and $|c_{\alpha\beta}|-\Phi_{\alpha\beta}$ planes (right panels) for HE-DUNE. The red curves in the bottom panels depict the effect of the addition of the $\nu_\tau$ appearance channel in the analysis. }
    \label{fig:contLIV}
\end{figure}

In Fig.~\ref{fig:contLIV} we show the allowed $1,2,3$ $\sigma$ contours in the $a_{\alpha\beta}-\Phi_{\alpha\beta}$ (top) and $c_{\alpha\beta}-\Phi_{\alpha\beta}$ planes (bottom). All the analysis have been performed fitting the data obtained using the SM parameters scanning over the LIV parameters one at-a-time. When computing the $\Delta\chi^2$, defined as:
\begin{equation}
    \Delta\chi^2=\chi^2(a_{\alpha\beta}/c_{\alpha\beta}\neq 0)-\chi^2(a_{\alpha\beta}/c_{\alpha\beta}= 0)\,,
\end{equation}
we fix all the not-showed LIV parameters to zero and we marginalize in the fit on all the oscillation parameters but the solar ones; we use Gaussian priors with the $1\sigma$ uncertainties\footnote{In our analysis, we assume that the octant of $\theta_{23}$ is known; in particular, as suggested by global fits in \cite{Esteban:2020cvm,nufit}, to lie in the lower octant.} as summarized in Tab. \ref{tab:t1} for $\theta_{23}$, $\theta_{13}$ and $\Delta m_{31}^2$ while leaving $\delta_{CP}$ free to vary in all its $[0,2\pi)^\circ$ range. 
The red lines in the bottom plots corresponding to $a_{\mu\tau}\, \,(c_{\mu\tau})$ depict the effect of the inclusion in the analysis of the $\nu_\tau$ appearance channel. We have verified that, as expected, in all other LIV parameters the effect  of the  $\nu_\tau$ appearance sample is marginal, a situation very similar to what discussed   in \cite{Ghoshal:2019pab,DeGouvea:2019kea}, where a non-negligible impact of the $\nu_\tau$ events were observed in the $\mu-\tau$ sector only in the context of propagation NSI.\raggedbottom
The 95\% limits on the moduli of the LIV parameters, obtained after marginalizing over the corresponding phases, are summarized in Tab.~\ref{tab:LIVbounds}. 
We observe that, fixing the octant and thus neglecting the degenerate LIV solutions found in \cite{Agarwalla:2023wft}, the HE-DUNE limits on the $|a_{e\beta}|$ parameters are weaker than the standard DUNE ones. This mainly comes from the fact that the $\nu_e$ appearance probability (strongly affected by $|a_{e\beta}|$) and consequently the number of $\nu_e$ events is larger at the first oscillation maximum than at the HE-DUNE energies. On the other hand, the limit on $|a_{\mu\tau}|$ results to be one order of magnitude more stringent in the HE-DUNE case due to the larger number of $\nu_\mu$ disappearance events. However, notice that this oscillation channel is strongly affected by the atmospheric mixing angle and by matter effects \cite{Agarwalla:2022xdo}; thus, the different procedure used in the LIV analysis in Ref. \cite{Agarwalla:2023wft}, where $\theta_{23}$ has been left free to vary in its 3$\sigma$ allowed range, might have amplified the differences between their standard DUNE and our HE-DUNE results. In fact, we checked that using our same minimization procedure, the standard DUNE bound on $|a_{\mu\tau}|$ would only be a factor 2 worse than the HE-DUNE one. The interplay between the magnitude of the LIV parameters and their phases is mostly visible in the case of $a_{\mu\tau}$ since the $\nu_\mu$ disappearance probability is directly proportional to $|a_{\mu\tau}|\cos\Phi_{\mu\tau}$ at the leading order. For this reason, the sensitivity is substantially worse when $\cos\Phi_{\mu\tau}\sim0$. When considering the other two parameters $a_{e\beta}$, the strong correlations between the new physics phase and the standard phase $\delta_{CP}$ makes the intepretation of the results as a  function of $\Phi_{\alpha\beta}$ less clear. However, it can be seen that the limits on $|a_{e\beta}|$ are only marginally impacted by the value of the LIV phase.  \\ 
As for the HE-DUNE limits on the CPT-even $c_{\alpha\beta}$ LIV parameters, as expected they are  better than the standard DUNE ones since their effects are amplified by the neutrino energy; the only exception is given by $c_{e\tau}$ which, for $\Phi_{e\tau}\sim90^\circ$, experiences a worsening in the sensitivity; indeed, as it can be seen from Eq.~\ref{eq:PMUE} and Fig.~\ref{fig:probLIV}, the probabilities when $\Phi_{e\tau}\sim90^\circ$ are very close to standard ones. Also in this case, except for  $c_{\mu\tau}$ (see eq. \ref{eq:MUMU}), the role of the LIV phase on the $|c_{\alpha\beta}|$ limits cannot be easily understood from analytical formulae, but our numerical results show that the limits on $|c_{e\mu}|$ and $|c_{e\tau}|$ do not drastically depend on the new $\Phi_{\alpha\beta}$ phase value. 

\begin{table}[]
\centering
\begin{tabular}{l|c|l|c|}
\cline{2-2} \cline{4-4}
                                    & \multicolumn{1}{l|}{95\% CL limit  ($\times 10^{-24} \,\, \rm{GeV}$)} &               & \multicolumn{1}{l|}{95\% CL limit ($\times 10^{-24}$)} \\ \hline
\multicolumn{1}{|l|}{$a_{e\mu}$}    & $<5.1$                                                                 & $c_{e\mu}$    & $<0.43$                                                \\ \hline
\multicolumn{1}{|l|}{$a_{e\tau}$}   & $<9.3$                                                                 & $c_{e\tau}$   & $<2.23$                                                \\ \hline
\multicolumn{1}{|l|}{$a_{\mu\tau}$} & $<1.12$ ($<1.0$)                                                       & $c_{\mu\tau}$ & $<0.66$ ($<0.64$)                                      \\ \hline
\end{tabular}
\caption{\label{tab:LIVbounds} 95\% bounds on the LIV parameters obtained for HE-DUNE. The upper limits have been obtained marginalizing over the LIV phases. The numbers in brackets refer to the foreseen improvement due to the addition of the $\nu_\tau$ appearance channel in the analysis.}
\end{table}

\section{The Long-Range Forces case }
\label{sec:LRFfull}

As it is well known, the neutrino flavor transition can be strongly affected by the presence of a matter medium, which can induce an effective potential  modifying the interaction Hamiltonian. However, the presence of BSM interactions between neutrinos and ordinary matter particles can in principle modify the matter potential term in the neutrino Hamiltonian. This is the case, for instance, of the widely studied vector and scalar Non-Standard Interactions (NSI) \cite{Denton:2020uda,deGouvea:2015ndi,Dev:2019anc,Bakhti:2020fde,Giarnetti:2021wur,Ge:2018uhz,Denton:2022pxt,ESSnuSB:2023lbg,Gupta:2023wct,Sarker:2023qzp}. Another interesting and less studied example is provided by the Long Range Forces (LRF) \cite{Grifols:1993rs,Grifols:1996fk,Grifols:2003gy,Mishra:2024riq,Agarwalla:2024ylc,Agarwalla:2023sng,Chatterjee:2015gta,Coloma:2020gfv}, which will be described in details in this section.

\subsection{Theoretical Framework}
The SM gauge group can be extended by anomaly free combination of the $U(1)$ symmetries $L_e$, $L_\mu$, $L_\tau$ and $B$ associated to lepton family number and baryon number. These combinations can be, for instance, $L_e-L_\mu$, $L_e-L_\tau$ and $L_\mu-L_\tau$ \cite{PhysRevD.43.R22,PhysRevD.44.2118,Foot:1994vd}\footnote{In \cite{Asai:2018ocx,Asai:2017ryy,Lou:2024fvw} it has been shown that $L_\alpha-L_\beta$ gauge symmetries can predict viable neutrino masses and mixing with the addition of Higgs-like particles charged under the new symmetries.}. Other combination have been discussed in the context of neutrino oscillation in \cite{Agarwalla:2024ylc}. The gauge boson of these symmetries is a massive neutral vector $Z^\prime$ which can mediate new physics interactions between neutrinos and matter. In the case of a large mediator mass, it is possible to study its effect in an Effective Field Theory (EFT) approach, which results on neutrino propagation affected in a vector NSI fashion. If, on the other hand, the mediator is extremely light, the flavor-dependent interaction forces between neutrinos and matter might become important over large distances. Given the huge interaction distance, proportional to $\lambda\sim m_{Z^\prime}^{-1}$, the matter potential term affecting neutrino oscillation will depend on the matter contained within a radius $\lambda$. \\
Let us first discuss the new interactions arising from these additional symmetries. In addition to the interactions mediated by the SM Z boson, for a $L_\alpha-L_\beta$ symmetry new lagrangian terms for the $Z^\prime$-induced interactions can be written as:
\begin{equation}
    \mathcal{L}_{Z^\prime}= g^\prime_{\alpha\beta }Z^\prime_\mu (\bar l_\alpha \gamma^\mu l_\alpha - \bar l_\beta \gamma^\mu l_\beta + \bar\nu_\alpha \gamma^\mu P_L \nu_\alpha-\bar\nu_\beta\gamma^\mu P_L \nu_\beta)\,,
\end{equation}
where $\nu_\alpha$ and $l_\alpha$ are the neutrino and the charged lepton fields and $P_L$ is the left-handed projection operator. There also exist a $Z-Z^\prime$ mixed term $\mathcal{L}_{\rm{mix}}$ which can introduce new four-fermion interactions proportional to the coupling $g'_{\alpha\beta}(\xi-\sin\theta_W \chi)$, where $\xi$ is the rotation angle between the gauge bosons eigenstates, $\chi$ is their kinetic mixing angle and $\theta_W$ is the usual Weinberg angle \cite{Agarwalla:2023sng}. This term allows not only neutrino-lepton new interactions, but also new contributions to the neutrino-nucleon scattering. However, since at large distances $\mathcal{L}_{Z^\prime}\gg\mathcal{L}_{\rm{mix}}$, the mixed term is important only in the $L_{\mu}-L_{\tau}$ case, for which neutrino-electron scattering mediated by $Z^\prime$ is prohibited. 

All these new interactions involving $Z^\prime$ induce a Yukawa-like potential coming from electrons and neutrons in the Universe which can affect neutrino oscillations \cite{Bustamante:2018mzu,Wise:2018rnb}. For a neutrino at a distance $d$ from a source of a number $N_e$ of electrons and for an $L_e-L_\beta$ symmetry, it can be written as \cite{Singh:2023nek,Agarwalla:2023sng,Mishra:2024riq}:
\begin{equation}
    V_{e\beta}=G^2_{e\beta} \frac{N_e}{4\pi d}e^{-m_{Z^\prime}d} \,.
    \label{eq:Veb}
\end{equation}
Under the $L_\mu-L_\tau$ symmetry, instead, the LRF effect comes from the mixing between $Z^\prime$ and the SM $Z$ boson. Considering the Universe to be neutral, the net potential in this case is only due to a $N_n$ number of neutrons, which generate a potential of the form:
\begin{equation}
    V_{\mu\tau}=G^2_{\mu\tau} \frac{e}{\sin\theta_{W}\cos\theta_W}\frac{N_n}{4\pi d}e^{-m_{Z^\prime}d} \,,
    \label{eq:Vmt}
\end{equation}
where $e$ is the electric charge. The effective coupling $G_{\alpha\beta}$ is the equivalent of the coupling $g^\prime_{\alpha\beta }$ in the case of an $L_e-L_\beta$ symmetry while, for the $L_\mu-L_\tau$, it is related to the lagrangian parameters through the relation $G_{\mu\tau}=\sqrt{g^\prime_{\mu\tau}(\xi-\sin\theta_W \chi)}$.  \\
Let us now consider the effect of LRF on the neutrino oscillations. In general, as previously mentioned, the neutrino propagation Hamiltonian always contains the vacuum and the standard matter potential terms shown in Eq.~\ref{eq:stdnuH}. The effect of the potential $V_{\alpha\beta}$ is  to add a new contribution of different structure depending on the consider symmetry:
\begin{equation}
\label{effpot}
\begin{gathered}
    H_{LRF}=\begin{pmatrix}
        V_{e\mu} & 0 & 0 \\
        0 & -V_{e\mu} & 0 \\
        0 & 0 & 0
    \end{pmatrix} \quad \text{for $L_e-L_\mu$,} \quad H_{LRF}=\begin{pmatrix}
        V_{e\tau} & 0 & 0 \\
        0 & 0 & 0 \\
        0 & 0 & -V_{e\tau}
    \end{pmatrix} \quad \text{for $L_e-L_\tau$,} \\ H_{LRF}=\begin{pmatrix}
        0 & 0 & 0 \\
        0 & V_{\mu\tau} & 0 \\
        0 & 0 & -V_{\mu\tau}
    \end{pmatrix} \quad \text{for $L_\mu-L_\tau$.}
    \end{gathered}
\end{equation}
Notice that the matrices in Eq.(\ref{effpot}) are similar to the standard matter potential; the main difference between the usual MSW contribution and the LRF one is that the former is a contact potential due to a very massive mediator (the SM $Z$ boson) while the latter encodes the effect of distant electron and neutrons sources on neutrino propagation due to an extremely light mediator (the $Z^\prime$) with a very large interaction length. \\
The computation of the oscillation probabilities is very cumbersome. In fact, 
in order to have observable effects of the long range potentials, we need the quantity $2 E V_{\alpha\beta}$ to be comparable to the vacuum oscillation frequency ($\Delta m_{31}^2/2E$ for LBL experiments); this implies that $V_{\alpha\beta}\sim V_{CC}\sim10^{-14}$ eV, where $V_{CC}$ is the usual matter potential term; since both potentials must be included into the evaluation, this produces very lengthy expressions for the transition probabilities. An example of analytical treatment of LRF have been carried out in \cite{Chatterjee:2015gta,Khatun:2018lzs,Agarwalla:2021zfr,Mishra:2024riq}. Notice that, in principle, LRF probabilities can also be deduced from those computed in presence of diagonal NSI parameters \cite{Agarwalla:2021zfr,Mishra:2024riq, Wise:2018rnb}. From them, one can recognize that that there exists a particular value of the neutrino energy for which a resonance occurs; 
in the case of $L_e-L_\beta$ symmetries, and neglecting the solar mass difference contribution, this condition reads \cite{Chatterjee:2015gta,Khatun:2018lzs}:
\begin{equation}
    E_{\rm{res}}=\frac{\Delta m_{31}^2 \cos2\theta_{13}}{2V_{CC}+3V_{e\beta}}\,.
\end{equation}
This means that the matter resonance occurs in presence of LRF at lower energies with respect to the standard MSW case. As in the LIV case, LRF have been widely studied in the literature in the context of neutrino oscillation. Limits on the LRF potentials and on the effective coupling appearing in $V_{\alpha\beta}$ were obtained, for example, in \cite{Joshipura:2003jh,Coloma:2020gfv,Bandyopadhyay:2006uh,Gonzalez-Garcia:2013usa,Honda:2007wv,Agarwalla:2023wft,Farzan:2018pnk,IceCube-Gen2:2020qha,Mishra:2024riq,Singh:2023nek,Agarwalla:2024ylc}. Focusing on the DUNE performances, the Fermilab-based experiment might set the following 95\% CL limits on the LRF potentials \cite{Singh:2023nek}:
\begin{equation}
    \begin{split}
        V_{e\mu}&<1.9\phantom{0} \times 10^{-14} \, \, \rm{eV} \\
         V_{e\tau}&<1.3\phantom{0} \times 10^{-14} \, \, \rm{eV} \\
          V_{\mu\tau}&<0.82 \times 10^{-14} \, \, \rm{eV} \,.
    \end{split}
    \label{eq:DUNElimitsLRF}
\end{equation}
The strongest limit can be put on $V_{\mu\tau}$ because it strongly affects the disappearance channel, which has a huge statistics in DUNE. 

\subsection{HE-DUNE results}

\begin{figure}[H]
    \centering
    \includegraphics[width=0.99\textwidth]{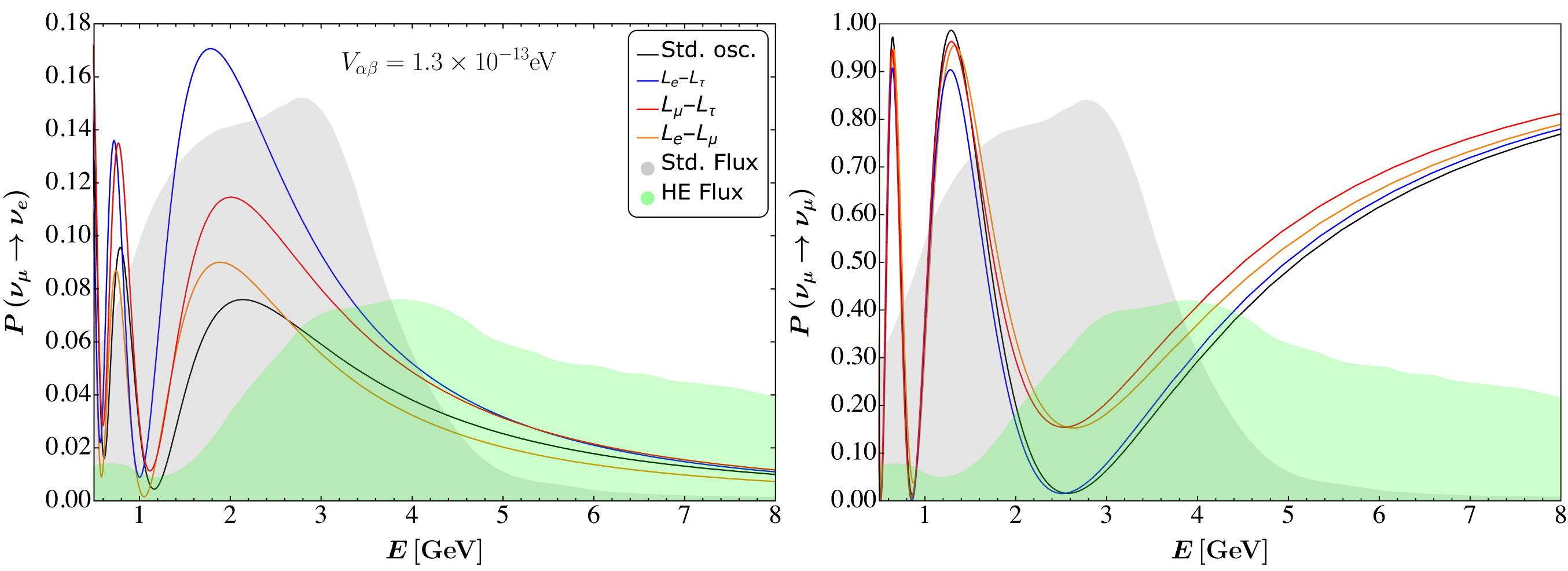}
    \caption{Same as Fig.~\ref{fig:probLIV}, but for the Long Range Force case. Left (right) plot shows the $\nu_e$ appearance ($\nu_\mu$ disappearance) probability. The blue, red and orange curves refers to the $L_e-L_\tau$, $L_\mu-L_\tau$ and $L_e-L_\mu$ cases, respectively. The potentials $V_{\alpha\beta}$ have been fixed to $1.3 \times 10^{-13}$ eV.
    }
    \label{fig:probLRF}
\end{figure}

\begin{figure}[H]
    \centering
    \includegraphics[width=0.75\textwidth]{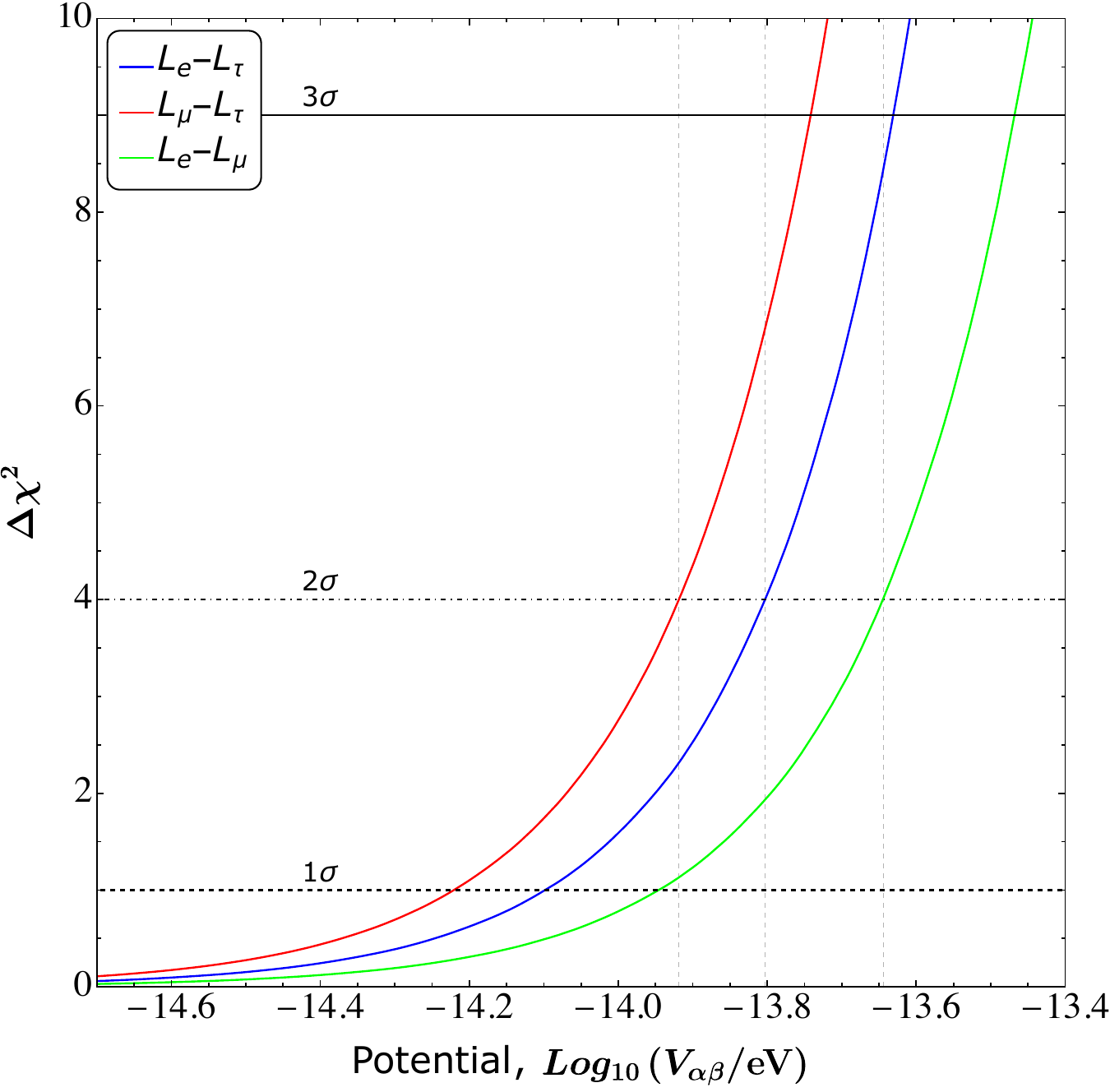}
    \caption{HE-DUNE sensitivity to the LRF potentials. Blue, red and green lines correspond to the $L_e-L_\tau$, $L_\mu-L_\tau$ and $L_e-L_\mu$ cases, respectively.}
    \label{fig:corrLRF}
\end{figure}

In this section we will explore the capabilities of HE-DUNE to constrain the LRF potential and the limits it might set on the strength of the new forces as well as on the new mediator mass. First, in Fig.~\ref{fig:probLRF} we plot the effect of Long-Range potential on the appearance and disappearance probabilities at the DUNE baseline.
The LRF potentials $V_{\alpha\beta}$ have been set to $1.3\times10^{-13}$ eV to show their effects when the LRF are of the same order of magnitude of the standard matter potential. It is clear that the appearance probability is enhanced at the first oscillation maximum for all three cases due to the LRF potential-induced resonances. In particular, $V_{e\tau}$ has the strongest effect while $V_{e\mu}$ has the mildest one. At higher energies, important for HE-DUNE, we observe that the $V_{\mu\tau}$ decreases the appearance probability while $V_{e\tau}$ and $V_{e\mu}$ increase it. On the other hand, the disappearance probability is enhanced at its first minimum, with $V_{\mu\tau}$ having the biggest impact for energies above 2.5 GeV. Notice that $V_{e\tau}$ has only a negligible effect on the disappearance probability. \\
Using the procedure described in Sec.~\ref{sec:LIV}, we estimated the bounds that HE-DUNE might be able to set on the LRF potentials. In Fig.~\ref{fig:corrLRF} we show the sensitivity to the three $V_{\alpha\beta}$ as obtained from $\Delta\chi^2$ as:
\begin{equation}
    \Delta\chi^2=\chi^2(V_{\alpha\beta}\neq0)-\chi^2(V_{\alpha\beta}=0) \,.
\end{equation}
We summarize the HE-DUNE 95\% CL limits in Tab.~\ref{tab:limitsLRF}. Comparing them with those in Eq.~\ref{eq:DUNElimitsLRF}, we observe that the HE-DUNE could set bounds on $V_{e\beta}$ and $V_{\mu\tau}$ which are 20\% and 35\% weaker than the DUNE ones, respectively. The reason is that, even though the effect of matter potentials are in general increased for large neutrino energies, long range potentials cause low-energy resonances in neutrino oscillation probabilities which can be probed better at the standard DUNE energies. 

\begin{table}[]
\centering
\begin{tabular}{c|c|c|c|}
\cline{2-4}
                                    & $V_{e\mu}$              & $V_{e\tau}$              & $V_{\mu\tau}$            \\ \hline
\multicolumn{1}{|l|}{95\% CL limit} & $<2.4\times10^{-14}$ eV & $<1.58\times10^{-14}$ eV & $<1.23\times10^{-14}$ eV \\ \hline
\end{tabular}
\caption{\label{tab:limitsLRF} 95\% CL limits on the Long Range Forces potentials obtained by HE-DUNE. }
\end{table}

\begin{figure}[H]
    \centering
    \includegraphics[width=0.99\textwidth]{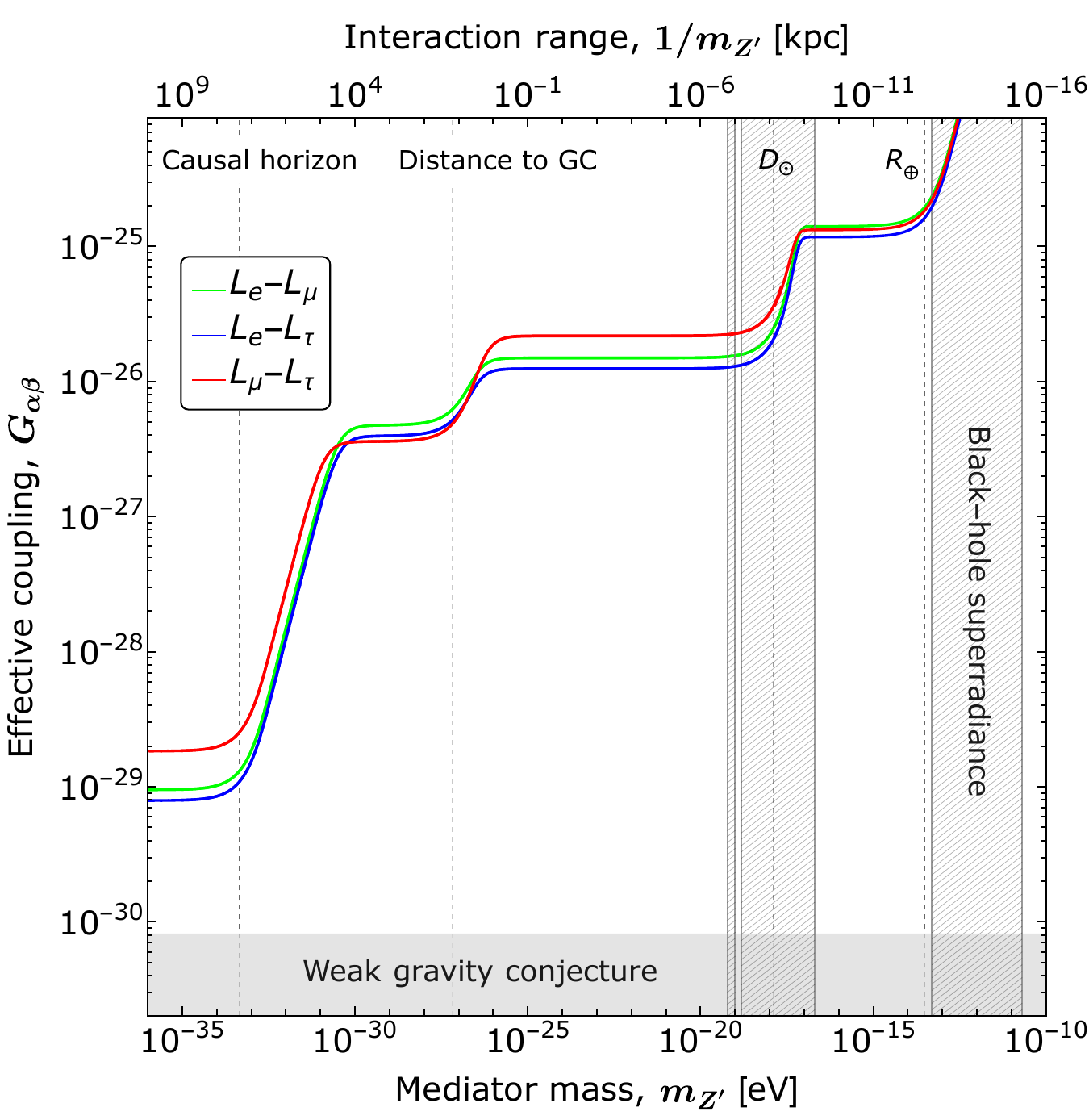}
    \caption{95\% CL excluded regions in the $m_{Z^\prime}-G_{\alpha\beta}$ plane, fixing the LRF potentials to the 95\% CL HE-DUNE limits showed in Tab.~\ref{tab:limitsLRF}. See text for details.
    }
    \label{fig:mvsG}
\end{figure}

One might use Eqs.~\ref{eq:Veb}~-~\ref{eq:Vmt} to find constraints to both the effective $Z^\prime$ coupling and its mass. Following Refs.~\cite{Singh:2023nek,Bustamante:2018mzu}, if we want to consider all the matter content of the Universe, we need to take into account neutrinos from matter sources away up to $10^3$ Gp, which corresponds to a $Z^\prime$ mediator mass in the range $10^{-10}-10^{-35}$ eV. Thus, we are dealing with an effective potential whose most important contributions are:
\begin{equation}
\label{fullpot}
    V_{\alpha\beta}=(V_{\alpha\beta})_{\rm Earth}+(V_{\alpha\beta})_{\rm Moon}+(V_{\alpha\beta})_{\rm Sun}+(V_{\alpha\beta})_{\rm Milky \, Way}+(V_{\alpha\beta})_{\rm Cosmol}\,.
\end{equation}
In Eq.(\ref{fullpot}), we consider the Moon and the Sun as point-like electron and neutron sources, with $(N_e)_{\rm Moon}=(N_n)_{\rm Moon}\sim5\times 10^{49}$ and $(N_e)_{\rm Sun}\sim4(N_n)_{\rm Sun}\sim 10^{57}$ \cite{Singh:2023nek}. On the other hand, we modelled the Earth as a continuous distribution with the same average density such as $(N_e)_{\rm Earth}=(N_n)_{\rm Earth}\sim4\times 10^{51}$; for the Milky Way, we divided the matter content in a thin disk, a thick disk, a central bulge
and a diffuse gas, following the reasonings on Refs. \cite{McMillan_2011,Miller_2013,Bustamante:2018mzu}, with $(N_e)_{\rm Milky \, Way}=(N_n)_{\rm Milky \, Way}\sim 10^{67}$. Finally, the cosmological matter has been included computing the whole potential described in \cite{Bustamante:2018mzu} at redshift z=0 as suggested in \cite{Singh:2023nek}; the total number of electrons and neutrons in this case has been fixed to 
$(N_e)_{\rm Cosmol}\sim10 (N_n)_{\rm Cosmol}\sim 10^{79}$. Once all the contributions to the long range potential are estimated, one can constrain both $m_{Z^\prime}$ and $G_{\alpha\beta}$ using the limits on $V_{\alpha\beta}$ in Tab.~\ref{tab:limitsLRF}. Our results are shown in Fig.~\ref{fig:mvsG} for the three cases $L_e-L_{\mu}$ (green line), $L_e-L_\tau$ (blue line) and $L_\mu-L_\tau$ (red line). In the upper part of the plot, we show the interaction length $1/m_{Z^\prime}$ corresponding to the a given mediator mass. 
The grey vertical bands show the parameters space excluded by two phenomena~\cite{Singh:2023nek}: black-hole superradiance and weak gravity conjecture. The former is related to the superradiant growth of an accumulation of very light vector bosons around extremely massive and gravitational bounded objects like supermassive black holes~\cite{Baryakhtar_2017}. The latter is related to a lower limit which might be set on the coupling in theories containing both gravity as weakest force and a $U(1)$ gauge interaction~\cite{Arkani_Hamed_2007}. 

We can observe that, at the specific distance at which the electron and neutron biggest sources are located, the limits on the effective couplings get stronger. In particular, for mediator masses lower than $10^{-33}$ eV, which correspond to roughly 10 Gpc (where the causal horizon is located), the bounds on $G_{\alpha\beta}$ get as low as $8\times10^{-30}$ for $G_{e\tau}$, $9.5\times10^{-30}$ for $G_{e\mu}$ and $2\times10^{-29}$ for $G_{\mu\tau}$. The latter is the weakest limit since it depends on $N_n$ which is smaller than $N_e$ for cosmological and solar matter, despite corresponding to $V_{\mu\tau}$ which is the potential bounded the most by HE-DUNE.

\section{The Large Extra Dimensions case }
\label{sec:LEDfull}

Neutrino oscillation are not predicted in the original version of the SM since the Higgs mechanism is not capable of providing non-zero neutrino masses and their smallness, 
compared to the other fermions, are difficult to contemplate in a theoretical general framework. These difficulties can be overcome by several BSM models \cite{deGouvea:2016qpx,King_2003}; among them, the Large Extra Dimensions (LED) theory \cite{Arkani-Hamed:1998wuz,Dienes_1999,Dvali_1999,Mohapatra_2001,Barbieri_2000,Davoudiasl_2002} not only provides a viable framework but also supply  an explanation for the enormous difference between  the Electroweak and the Planck scale \cite{Arkani_Hamed_1998_1,Arkani_Hamed_1999}. The main idea is to introduce sterile right-handed neutrino fields, which are singlet under the SM group but propagate in a $4+N_{ED}$ dimensional space-time, where $N_{ED}$ is the number of space-like extra dimensions. Usual Yukawa lagrangian terms can then be built using left handed neutrinos and Higgs fields which live in the usual 4 dimensional space-time. However, the masses coming from such terms are heavily suppressed with respect to  other fermion  masses due to the much smaller wave function normalizations in the large volume of the extra dimensions. In the following we will explore the phenomenological implications of the existence of LED on neutrino oscillation.

\subsection{Theoretical Framework}

Following the approach proposed for neutrino oscillation studies in the context of LED\cite{Esmaili_2014,Machado_2011,Machado_2012,Basto_Gonzalez_2013,Girardi_2014,Rodejohann_2014,Carena:2017qhd,Stenico_2018,Arg_elles_2020,bastogonzalez2022shortbaseline,Forero_2022,Berryman_2016}, we will focus on $N_{ED}=1$; this single LED is compactified on a circle of radius $R_{ED}$ and, for the sterile neuttrinos, this gives rise to Kaluza-Klein (KK) modes in the 4 dimensional space-time. It is also possible to  consider the presence of more extra dimensions, whose compatification radius is much smaller than $R_{ED}$ without changing the effect of LED on the neutrino oscillation. 
The model is built adding three massless five-dimensional fermion fields $\Psi^\alpha=(\psi_L^\alpha,\psi_R^\alpha)$ to the SM. After the compactification of the fifth dimension with the proper boundary conditions, the $\Psi^\alpha$ fields appear in the 4-dimensional space time as an infinite tower of KK states $\psi^{n}$, where $n$ is any integer number. Identifying the zero mode as the right-handed neutrinos $\nu_R^{\alpha}=\psi_R^{\alpha(0)}$ and writing $\nu_{L,R}^{\alpha(n)}=(\psi_{L,R}^{\alpha(-n)}+\psi_{L,R}^{\alpha(n)})/\sqrt{2}$ 
\cite{Berryman_2016}, the neutrino mass term is \cite{Berryman_2016,Barbieri_2000,Dienes_1999,Arkani_Hamed_1998_1}:
\begin{equation}
    \mathcal{L}_{\rm{LED}}=m_{\alpha\beta}^D \left(\bar\nu_R^\alpha\nu_L^\beta+\sqrt{2}\sum_{n=1}^\infty \bar\nu_{R}^{\alpha(n)}\nu_L^\beta\right)+\sum_{n=1}^\infty \frac{n}{R_{ED}}\bar\nu_{R}^{\alpha(n)}\nu_L^{\alpha(n)}+h.c.\,,
    \label{eq:LEDlag}
\end{equation}
where $m^D_{\alpha\beta}$ is the Dirac mass matrix. Rewriting the mass eigenstates as $N^i=\left(\nu^{i(0)},\nu^{i(1)},....\right)^T$, then $\mathcal{L}_{\rm{LED}}$ can be written as $\sum_{i=1}^3 \bar N_R^i M^i N_L^i$ where the infinite mass matrices is:
\begin{equation}
    M_i=\begin{pmatrix}
        m_i & \sqrt{2}m_i & \sqrt{2}m_i & \dots & \sqrt{2}m_i & \dots \\
        0 & 1/R_{ED} & 0 &\dots & 0 &\dots \\
        0 & 0 & 2/R_{ED} & \dots & 0 &\dots \\
        0 & 0 & 0& \dots & n/R_{ED} & \dots \\
        \vdots & \vdots &\vdots &\vdots & \vdots &\ddots
    \end{pmatrix}\,,
    \label{eq:mmatrixLED}
\end{equation}
in which the $m_i$ are the eigenvalues of the neutrino Dirac mass matrix $m_{\alpha\beta}^D$ in Eq.~\ref{eq:LEDlag}. The neutrino mixing among the active states is then defined by the equation \cite{Forero_2022,Berryman_2016}
\begin{equation}
    \nu_{\alpha}=\sum_{i=1}^3 U_{\alpha i} \sum_{n=0}^{\infty} V_{in} \nu_{i}^{(n)}\,,
\end{equation}
where $U$ is the PMNS mixing matrix and $V$ is the \textit{effective mixing matrix} among the KK excitations. Its elements can be written as:
\begin{equation}
    (V_{in})^2=\frac{2}{1+\pi^2 (m_i R_{ED})^2+(\lambda_i^{(n)})^2/(m_i R_{ED})^2}\,,
\end{equation}
with $\lambda_{i}^{(n)}$ being the eigenvalues of the matrices $R^2_{ED}M^\dagger_i M_i$ that can be found as solutions of the equation \cite{Dienes_1999,Machado_2011,Barbieri_2000,Berryman_2016,Forero_2022}
\begin{equation}
    \lambda_{i}^{(n)}-\pi (m_i R_{ED})^2\cot\left(\pi \lambda_{i}^{(n)}\right)=0 \,.
    \label{eq:trasceq}
\end{equation}
Notice that the masses of the KK states are in this case $m_i^{(n)}=\lambda_i^{(n)}/R_{ED}$; since the solutions of Eq.~\ref{eq:trasceq} satisfy the relation $n\leq \lambda_i^{(n)}<(n+1/2)$, we can roughly say that $m_i^{(n)}\sim n/R_{ED}$ \cite{Berryman_2016}. Once we fix the experimental observation of the mass splittings to be equal to the differences $\Delta m_{21}^2=[(\lambda_2^{(0)})^2-(\lambda_1^{(0)})^2]/R_{ED}^2$ and $\Delta m_{31}^2=[(\lambda_3^{(0)})^2-(\lambda_1^{(0)})^2]/R_{ED}^2$, we are left with the standard three neutrino mixing modified by the effect of the mixing between the active neutrinos and an infinite number of sterile neutrinos. The only non-standard parameters of the model are then the compactification radius $R_{ED}$ and the smallest Dirac mass $m_{1}$. The oscillation probabilities can be obtained, in vacuum, as
\begin{equation}
    P_{\alpha\beta}=\left| \sum_{i=1}^3 \sum_{n=0}^{\infty}U^*_{\alpha i}U_{\beta i}V_{in}^2 \, \mathrm{exp}\left(-i\frac{(m_i^{(n)})^2L}{2E}\right) \right|
\end{equation}
where $L$ is the baseline.
In the limit $m_i R_{ED}\to 0$, we observe that $m_i^{(n)}\to\infty$ for $n\neq0$ and $V_{in}\to\delta_{0n}$, making the oscillation phenomenology identical to the standard one. \\
It has been shown that, apart from the appearance of new matter resonances at high neutrino energies \cite{Esmaili_2014} ($E\gg 1 \, \, \rm{TeV}$), the LED phenomenology does not change significantly if we include more than 2 KK modes. Indeed, higher modes would imply larger masses $m_{i}^{(n)}$ and smaller matrix elements $V_{in}$ \cite{Forero_2022}. Thus, one might study neutrino oscillation in presence of LED including in the model only a limited number of KK modes. In this context, it is possible to treat the LED case with a number $n_{\rm {KK}}$ of KK modes as a $3+3 n_{\rm{KK}}$ sterile neutrino model \cite{Esmaili_2014,Berryman_2016}, where all the non-standard mixing angles and mass splittings can be written in terms of the two LED parameters\footnote{Notice that a general model with sterile neutrinos cannot be mimicked by a LED model in general; for instance, the presence of Large Extra Dimensions does not generate new sources of CP-violation, unlike the sterile neutrinos hypothesis \cite{Berryman_2016,Giarnetti:2021wur}.}. \\
The expression of the oscillation probabilities in the LED case are very cumbersome; however, in the limit $m_i R_{ED}\ll 1$, some expansions for the mass eigenstates and mixing matrix elements have been obtained in Ref.~\cite{Davoudiasl_2002,Forero_2022}:
\begin{equation}
    \begin{split}
        &m_i^{(0)}=m_i\left[1-\frac{\pi^2}{6}(m_i R_{ED})^2+\dots\right]\sim m_i \\
        &m_i^{(n)}=\frac{n}{R_{ED}}\left[1+\frac{(m_i R_{ED})^2}{n^2}+\dots\right]\sim \frac{n}{R_{ED}}\\
        &V_{i0}=1-\frac{\pi^2}{6}(m_i R_{ED})^2+\dots\sim1 \\
         &V_{in}=\sqrt{2}\frac{m_i R_{ED}}{n}\left[1-\frac{3}{2}\frac{(m_i R_{ED})^2}{n^2}+\dots\right]\sim \sqrt{2}\frac{m_i R_{ED}}{n}
    \end{split}
    \label{eq:expansionsVm}
\end{equation}
from which it is clear that, as already mentioned, the corrections to the standard oscillation case become negligible as $n$ increases. \\
The complete picture of neutrino oscillation in presence of LED is obtained when matter effects are added. Then, the oscillation probabilities can be obtained solving the Schroedinger-like evolution equation \cite{Berryman_2016}
\begin{equation}
    i\frac{d}{dr}N_{i}=\frac{1}{2E}M_i^\dagger M_i N_i +\sum_{j=1}^3 \lim_{n\to\infty}\begin{pmatrix}
        \rho_{ij} & 0_{1\times n} \\ 0_{n\times 1} & 0_{n\times n}\,,
    \end{pmatrix}N_j\,,
\end{equation}
where the $N_i$ infinite vector of neutrino states and $M_i$ matrices have already been defined in Eq. \ref{eq:mmatrixLED} and above, while the quantity $\rho_{ij}$ is defined as
\begin{equation}
    \rho_{ij}=\sum_{\alpha}U_{\alpha i}^* U_{\alpha j}(\delta_{\alpha e}V_{CC}-V_{NC})\,,
\end{equation}
with $U$ being the standard $3\times3$ PMNS matrix, $V_{CC}=\sqrt{2}G_F n_e$ the usual matter potential and $V_{NC}=-2\sqrt{2}G_F n_n$ the neutral current matter potential that can no longer be neglected due to the presence of sterile states. 

The limits that current experiments could set on the parameter space in the LED case have been discussed, for instance,  in Ref.~\cite{Forero_2022}. Notice also that the short baseline, reactor and gallium anomalies, which have been explained with the presence of light sterile neutrinos, could be explained in presence of LED~\cite{Carena:2017qhd,Berryman_2016,Forero_2022}. In the context of the future DUNE experiment, the expected performances of the experiment have been explored in details in Ref.~\cite{Berryman_2016}. The upper limits on the $R_{ED}$ parameter depend on the lightest neutrino mass value; in particular, in the scenario when $m_1\to0$ eV and the probabilities only depend on $R_{ED}$, we expect for DUNE at 95\% CL:
\begin{equation}
    \begin{split}
        R_{ED}&<0.32 \, \, \mathrm{\mu m} \,,
    \end{split}
\end{equation}
while for $m_1\sim0.05$ eV, we expect 
\begin{equation}
    \begin{split}
        R_{ED}&<0.22 \, \, \mathrm{\mu m} \,.
    \end{split}
    \label{eq:limitDUNE0.05EV}
\end{equation}
On the other hand, it is impossible to set an absolute limit on $m_1$ unless data are generated considering $R_{ED}\neq0$; in that case DUNE might set a lower limit on the absolute neutrino mass \cite{Berryman_2016}.\footnote{In \cite{Arguelles:2022tki}, a DUNE analysis has been repeated using a different experimental configuration; they found weaker limits on $R_{ED}$.}

\subsection{HE-DUNE results}

\begin{figure}[H]
    \centering
    \includegraphics[width=0.89\textwidth]{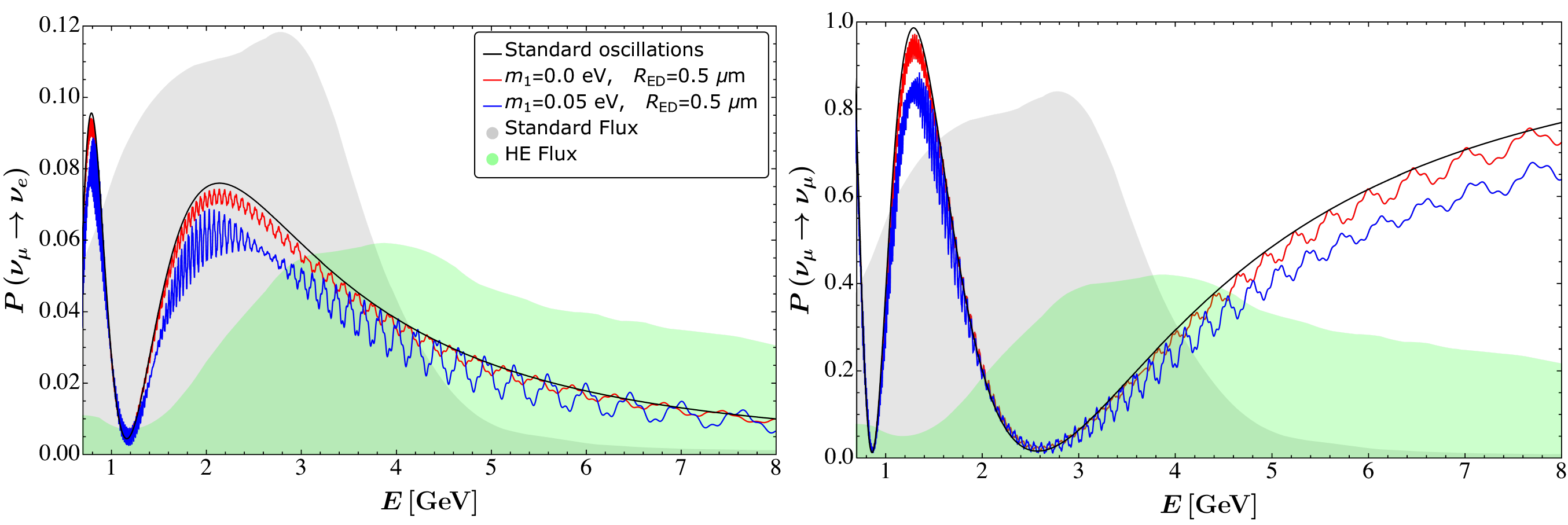}
    \caption{Same as Fig.~\ref{fig:probLIV} but in the Large Extra Dimension case. Red (blue) curves have been obtained fixing $R_{ED}=0.5$ $\mu m$ and $m_1=0.0$ eV ($m_1=0.05$ eV).
    }
    \label{fig:probLED}
\end{figure}

\begin{figure}[H]
    \centering
    \includegraphics[width=0.89\textwidth]{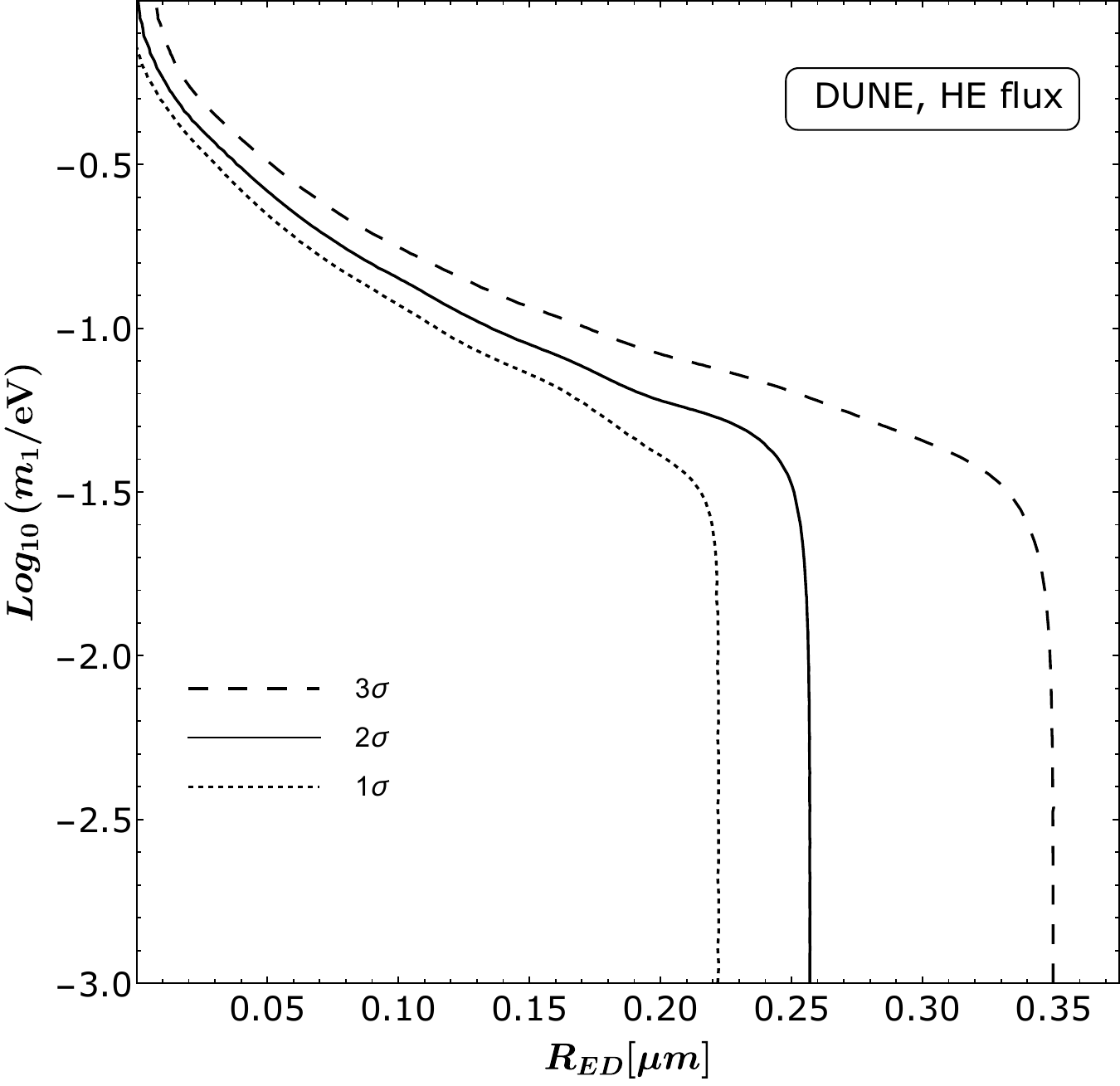}
    \caption{$1\sigma$ (dotted) $2\sigma$ (solid) and $3\sigma$ (dashed) allowed regions in the $R_{ED}-m_1$ plane for HE-DUNE.
    }
    \label{fig:boundLED}
\end{figure}

In this section we discuss the effects of LED at HE-DUNE. For the $\nu_e$ appearance and $\nu_\mu$ disappearance probabilities (see Fig.~\ref{fig:probLED}), we consider as reference two possible lightest neutrino masses $m_1$, namely $0.0$ and $0.05$ eV. For the compactification radius, we choose $0.5 \, \mathrm{\mu m}$ which correspond to $R_{ED}^{-1}=0.38 \, \mathrm{eV^{-1}}$. In our computations we included 3 KK modes, even though we checked that our results are only negligibly affected by the inclusion of the second and the third KK modes. From Fig.~\ref{fig:probLED} we see that the main effect of LED is the occurrence of new fast oscillations driven by the large mass splittings between the active states and the heavy KK excitations. The amplitude of the oscillations depends on the values of the $V_{in}$ matrix elements. In addition,  the presence of LED decreases both the appearance and disappearance probabilities at the first oscillation maximum, since the value of $V_{i0}$ is always less than 1, see Eq.~\ref{eq:expansionsVm}. Moreover, at a fixed $R_{ED}$, the probabilities in general decrease as $m_1$ increases. 
The differences between probabilities around the first oscillation maximum (DUNE) and the high energy region (HE-DUNE) is mainly due to the fact that fast oscillation driven by KK becomes slower and with a larger amplitude when the neutrino energy increases. For this reason, with high-energy neutrinos it might be possible to resolve these oscillations allowing for better constraints on the LED model \cite{Arguelles:2022tki}.\\ 
In Fig.~\ref{fig:boundLED} we show the allowed parameters space in the $R_{ED}-m_1$ plane in the HE-DUNE case. The analysis have been performed using the procedure described in the previous two sections. Here,  we can see that at $2\sigma$ the weakest limit for $R_{ED}$ reached for $m_1\to0$ is $0.258$ $\mathrm{\mu m}$. This constraint is better than the standard DUNE one. As noted above, one of the main reasons for that is the possibility to recognize fast oscillations at higher energies, where they occur with a smaller frequency with respect to the lower energy region probed by standard DUNE. Another interesting feature is that for $m_1<0.04$ eV, HE-DUNE constraint on $R_{ED}$ becomes independent on the lightest neutrino mass. This means that if $m_1$ is, for instance, $0.05$ eV, then the standard DUNE experiment outperforms HE-DUNE (see Eq.~\ref{eq:limitDUNE0.05EV}).

\section{Conclusions}
\label{sec:concl}

Neutrino oscillation are the most well established phenomenon beyond the Standard Model of particle physics. Despite several experiments have been able to measure the oscillation parameters with a few percents uncertainty, there are still some unknowns like the neutrino mass hierarchy and the amount of leptonic CP violation (if any). Moreover, oscillation searches are of a great interest since several new physics models can affect the neutrino propagation and thus modify the related probabilities. Future oscillation facilities are expected to reach a great precision on the measurements of the mixing parameters; at the same time, they could provide a great probe for new physics models involving neutrinos. In this work, we considered the capabilities of the High-Energy flux configuration of the future DUNE experiment (HE-DUNE). The possibility to employ this broad neutrino flux, which might reach more than 15 GeV in energy, has been envisaged in order to have access to $\nu_\mu\to\nu_\tau$ oscillations, which could not be easily observed with the standard DUNE configuration due to the energy threshold of CC $\nu_\tau$ interactions. In addition, HE-DUNE could in principle also be very useful to constrain new physics scenarios where the non-standard oscillation effects are more pronounced at high energies. \\
In this work we considered three different new physics models:
\begin{itemize}
    \item \textbf{Lorentz Invariance Violation (LIV)}: in this model, the neutrino lagrangian density is modified trough several Lorentz Violating operators, both CPT-even and CPT-odd. The presence of such operators modifies the neutrino propagation Hamiltonian with the addition of two hermitian matrices $a_{\alpha\beta}$ (CPT-violating) and $c_{\alpha\beta}$ (CPT-conserving). The effects of the second matrix increase linearly with the neutrino energy. We studied in Sec.~\ref{sec:LIV} the sensitivity of HE-DUNE to the off-diagonal LIV parameters. We found that the limits on the moduli of CPT-violating parameters $|a_{\alpha\beta}|$ are worse than the ones that the standard DUNE is expected to set. On the other hand, HE-DUNE capabilities should exceed the standard DUNE ones in constraining energy-enhanced effects of CPT-conserving LIV parameters $|c_{\alpha\beta}|$.
    \item \textbf{Long Range Forces (LRF)}: in this model we expect that new interactions with an ultra-light mediator, with a very long interaction length, arise from a gauge $U(1)$ symmetry of the form $L_\alpha-L_\beta$. These interactions can modify the matter potential term in the neutrino oscillation Hamiltonian. We showed that the limits from HE-DUNE on this new potential are rather stringent but not enough to overcome the standard DUNE ones. We also computed the limits on the coupling of the new interaction as well as on the mass of the new mediator. These are correlated to the interaction length since, depending on that, neutrino might experience the potential generated from various astrophysical matter densities.
    \item \textbf{Large Extra Dimensions (LED)}: if right-handed neutrinos are singlets under the SM group but they can propagate in a space-time with more than 4 dimensions, the smallness of neutrino masses can be naturally explained. In the case in which one of the new dimensions is compactified in a sphere with a relatively large radius, the Kaluza-Klein excitations of the neutrino states can be treated as sterile neutrinos involved in the oscillation. In this approach, the transition probabilities depend not only on the standard mixing parameters, but also on the smallest Dirac neutrino mass and on the compactification radius $R_{ED}$ of the large extra dimension. We showed that the limit that HE-DUNE might set on $R_{ED}$,  for small enough lightest neutrino mass, is better than the standard DUNE one. This is because the fast active-sterile oscillations coming from the Kaluza-Klein states might be resolved better at high energies than at the lower ones.
\end{itemize}
In conclusion, the DUNE high-energy flux might be useful not only to collect a large sample of $\nu_\tau$ events but also to set stringent limits on new physics parameters. This suggests that a HE-DUNE run could provide several information on the BSM neutrino physics which are complementary to the standard DUNE ones. 



\bibliographystyle{JHEP}
\bibliography{main.bib}

\providecommand{\href}[2]{#2}\begingroup\raggedright\begin{thebibliography}{100}

\bibitem{Super-Kamiokande:1998kpq}
{\bf Super-Kamiokande} Collaboration, Y.~Fukuda et~al., {\it {Evidence for oscillation of atmospheric neutrinos}},  {\em Phys. Rev. Lett.} {\bf 81} (1998) 1562--1567, [\href{http://arxiv.org/abs/hep-ex/9807003}{{\tt hep-ex/9807003}}].

\bibitem{XU2023104043}
X.-J. Xu, Z.~Wang, and S.~Chen, {\it {Solar neutrino physics}},  {\em Prog. Part. Nucl. Phys.} {\bf 131} (2023) 104043, [\href{http://arxiv.org/abs/2209.14832}{{\tt arXiv:2209.14832}}].

\bibitem{CHOUBEY2016235}
S.~Choubey, {\it {Atmospheric neutrinos: Status and prospects}},  {\em Nucl. Phys. B} {\bf 908} (2016) 235--249, [\href{http://arxiv.org/abs/1603.06841}{{\tt arXiv:1603.06841}}].

\bibitem{Cao:2017drk}
L.~J. W.~J. Cao and Y.~F. Wang, {\it {Reactor Neutrino Experiments: Present and Future}},  {\em Ann. Rev. Nucl. Part. Sci.} {\bf 67} (2017) 183--211, [\href{http://arxiv.org/abs/1803.10162}{{\tt arXiv:1803.10162}}].

\bibitem{Mezzetto:2020jka}
M.~Mezzetto and F.~Terranova, {\it {Three-flavour oscillations with accelerator neutrino beams}},  {\em Universe} {\bf 6} (2020), no.~2 32, [\href{http://arxiv.org/abs/2002.04890}{{\tt arXiv:2002.04890}}].

\bibitem{KamLAND:2008dgz}
{\bf KamLAND} Collaboration, S.~Abe et~al., {\it {Precision Measurement of Neutrino Oscillation Parameters with KamLAND}},  {\em Phys. Rev. Lett.} {\bf 100} (2008) 221803, [\href{http://arxiv.org/abs/0801.4589}{{\tt arXiv:0801.4589}}].

\bibitem{KamLAND:2010fvi}
{\bf KamLAND} Collaboration, A.~Gando et~al., {\it {Constraints on $\theta_{13}$ from A Three-Flavor Oscillation Analysis of Reactor Antineutrinos at KamLAND}},  {\em Phys. Rev. D} {\bf 83} (2011) 052002, [\href{http://arxiv.org/abs/1009.4771}{{\tt arXiv:1009.4771}}].

\bibitem{KamLAND:2013rgu}
{\bf KamLAND} Collaboration, A.~Gando et~al., {\it {Reactor On-Off Antineutrino Measurement with KamLAND}},  {\em Phys. Rev. D} {\bf 88} (2013), no.~3 033001, [\href{http://arxiv.org/abs/1303.4667}{{\tt arXiv:1303.4667}}].

\bibitem{DayaBay:2012fng}
{\bf Daya Bay} Collaboration, F.~P. An et~al., {\it {Observation of electron-antineutrino disappearance at Daya Bay}},  {\em Phys. Rev. Lett.} {\bf 108} (2012) 171803, [\href{http://arxiv.org/abs/1203.1669}{{\tt arXiv:1203.1669}}].

\bibitem{Esteban:2020cvm}
I.~Esteban, M.~C. Gonzalez-Garcia, M.~Maltoni, T.~Schwetz, and A.~Zhou, {\it {The fate of hints: updated global analysis of three-flavor neutrino oscillations}},  {\em JHEP} {\bf 09} (2020) 178, [\href{http://arxiv.org/abs/2007.14792}{{\tt arXiv:2007.14792}}].

\bibitem{Hyper-Kamiokande:2018ofw}
{\bf Hyper-Kamiokande} Collaboration, K.~Abe et~al., {\it {Hyper-Kamiokande Design Report}},  \href{http://arxiv.org/abs/1805.04163}{{\tt arXiv:1805.04163}}.

\bibitem{DUNE:2015lol}
{\bf DUNE} Collaboration, R.~Acciarri et~al., {\it {Long-Baseline Neutrino Facility (LBNF) and Deep Underground Neutrino Experiment (DUNE)}: {Conceptual Design Report, Volume 2: The Physics Program for DUNE at LBNF}},  \href{http://arxiv.org/abs/1512.06148}{{\tt arXiv:1512.06148}}.

\bibitem{Himmel:2020kct}
A.~Himmel, {\it {New Oscillation Results from the NOvA Experiment}}, .

\bibitem{Batkiewicz-Kwasniak:2022vrj}
{\bf T2K} Collaboration, M.~Batkiewicz-Kwasniak, {\it {The Latest T2K Neutrino Oscillation Results and the Future of the T2K and Hyper-Kamiokande Experiments}},  {\em Acta Phys. Polon. Supp.} {\bf 15} (2022), no.~3 23.

\bibitem{Rahaman:2021zzm}
U.~Rahaman and S.~Raut, {\it {On the tension between the latest NOvA and T2K data}},  \href{http://arxiv.org/abs/2112.13186}{{\tt arXiv:2112.13186}}.

\bibitem{Ghoshal:2019pab}
A.~Ghoshal, A.~Giarnetti, and D.~Meloni, {\it {On the role of the $\nu_{\tau}$ appearance in DUNE in constraining standard neutrino physics and beyond}},  {\em JHEP} {\bf 12} (2019) 126, [\href{http://arxiv.org/abs/1906.06212}{{\tt arXiv:1906.06212}}].

\bibitem{Masud:2017bcf}
M.~Masud, M.~Bishai, and P.~Mehta, {\it {Extricating New Physics Scenarios at DUNE with Higher Energy Beams}},  {\em Sci. Rep.} {\bf 9} (2019), no.~1 352, [\href{http://arxiv.org/abs/1704.08650}{{\tt arXiv:1704.08650}}].

\bibitem{DeGouvea:2019kea}
A.~De~Gouv\^ea, K.~J. Kelly, G.~V. Stenico, and P.~Pasquini, {\it {Physics with Beam Tau-Neutrino Appearance at DUNE}},  {\em Phys. Rev. D} {\bf 100} (2019), no.~1 016004, [\href{http://arxiv.org/abs/1904.07265}{{\tt arXiv:1904.07265}}].

\bibitem{DeRomeri:2023dht}
V.~De~Romeri, C.~Giunti, T.~Stuttard, and C.~A. Ternes, {\it {Neutrino oscillation bounds on quantum decoherence}},  {\em JHEP} {\bf 09} (2023) 097, [\href{http://arxiv.org/abs/2306.14699}{{\tt arXiv:2306.14699}}].

\bibitem{MammenAbraham:2022xoc}
R.~Mammen~Abraham et~al., {\it {Tau neutrinos in the next decade: from GeV to EeV}},  {\em J. Phys. G} {\bf 49} (2022), no.~11 110501, [\href{http://arxiv.org/abs/2203.05591}{{\tt arXiv:2203.05591}}].

\bibitem{DUNE:2016hlj}
{\bf DUNE} Collaboration, R.~Acciarri et~al., {\it {Long-Baseline Neutrino Facility (LBNF) and Deep Underground Neutrino Experiment (DUNE)}: {Conceptual Design Report, Volume 1: The LBNF and DUNE Projects}},  \href{http://arxiv.org/abs/1601.05471}{{\tt arXiv:1601.05471}}.

\bibitem{DUNE:2020jqi}
{\bf DUNE} Collaboration, B.~Abi et~al., {\it {Long-baseline neutrino oscillation physics potential of the DUNE experiment}},  {\em Eur. Phys. J. C} {\bf 80} (2020), no.~10 978, [\href{http://arxiv.org/abs/2006.16043}{{\tt arXiv:2006.16043}}].

\bibitem{DUNE:2020ypp}
{\bf DUNE} Collaboration, B.~Abi et~al., {\it {Deep Underground Neutrino Experiment (DUNE), Far Detector Technical Design Report, Volume II: DUNE Physics}},  \href{http://arxiv.org/abs/2002.03005}{{\tt arXiv:2002.03005}}.

\bibitem{DUNE:2021tad}
{\bf DUNE} Collaboration, V.~Hewes et~al., {\it {Deep Underground Neutrino Experiment (DUNE) Near Detector Conceptual Design Report}},  {\em Instruments} {\bf 5} (2021), no.~4 31, [\href{http://arxiv.org/abs/2103.13910}{{\tt arXiv:2103.13910}}].

\bibitem{HEDUNE-1}
{\it {http://home.fnal.gov/~ljf26/DUNEFluxes/}}, .

\bibitem{HEDUNE-2}
M.~Bishai and M.~Dolce, {\it {Optimization of the LBNF/DUNE beamline for tau neutrinos}}, .

\bibitem{HEDUNE-3}
{\it {https://glaucus.crc.nd.edu/DUNEFluxes/}}, .

\bibitem{DUNE:2016ymp}
{\bf DUNE} Collaboration, T.~Alion et~al., {\it {Experiment Simulation Configurations Used in DUNE CDR}},  \href{http://arxiv.org/abs/1606.09550}{{\tt arXiv:1606.09550}}.

\bibitem{DUNE:2021cuw}
{\bf DUNE} Collaboration, B.~Abi et~al., {\it {Experiment Simulation Configurations Approximating DUNE TDR}},  \href{http://arxiv.org/abs/2103.04797}{{\tt arXiv:2103.04797}}.

\bibitem{Colladay:1996iz}
D.~Colladay and V.~A. Kostelecky, {\it {CPT violation and the standard model}},  {\em Phys. Rev. D} {\bf 55} (1997) 6760--6774, [\href{http://arxiv.org/abs/hep-ph/9703464}{{\tt hep-ph/9703464}}].

\bibitem{Colladay:1998fq}
D.~Colladay and V.~A. Kostelecky, {\it {Lorentz violating extension of the standard model}},  {\em Phys. Rev. D} {\bf 58} (1998) 116002, [\href{http://arxiv.org/abs/hep-ph/9809521}{{\tt hep-ph/9809521}}].

\bibitem{Diaz:2011ia}
J.~S. Diaz and A.~Kostelecky, {\it {Lorentz- and CPT-violating models for neutrino oscillations}},  {\em Phys. Rev. D} {\bf 85} (2012) 016013, [\href{http://arxiv.org/abs/1108.1799}{{\tt arXiv:1108.1799}}].

\bibitem{Greenberg:2002uu}
O.~W. Greenberg, {\it {CPT violation implies violation of Lorentz invariance}},  {\em Phys. Rev. Lett.} {\bf 89} (2002) 231602, [\href{http://arxiv.org/abs/hep-ph/0201258}{{\tt hep-ph/0201258}}].

\bibitem{Kostelecky:1991ak}
V.~A. Kostelecky and R.~Potting, {\it {CPT and strings}},  {\em Nucl. Phys. B} {\bf 359} (1991) 545--570.

\bibitem{Kostelecky:1994rn}
V.~A. Kostelecky and R.~Potting, {\it {CPT, strings, and meson factories}},  {\em Phys. Rev. D} {\bf 51} (1995) 3923--3935, [\href{http://arxiv.org/abs/hep-ph/9501341}{{\tt hep-ph/9501341}}].

\bibitem{Kostelecky:1995qk}
V.~A. Kostelecky and R.~Potting, {\it {Expectation values, Lorentz invariance, and CPT in the open bosonic string}},  {\em Phys. Lett. B} {\bf 381} (1996) 89--96, [\href{http://arxiv.org/abs/hep-th/9605088}{{\tt hep-th/9605088}}].

\bibitem{Kostelecky:2003cr}
V.~A. Kostelecky and M.~Mewes, {\it {Lorentz and CPT violation in neutrinos}},  {\em Phys. Rev. D} {\bf 69} (2004) 016005, [\href{http://arxiv.org/abs/hep-ph/0309025}{{\tt hep-ph/0309025}}].

\bibitem{PhysRevD.39.683}
V.~A. Kosteleck\'y and S.~Samuel, {\it Spontaneous breaking of lorentz symmetry in string theory},  {\em Phys. Rev. D} {\bf 39} (Jan, 1989) 683--685.

\bibitem{PhysRevLett.63.224}
V.~A. Kosteleck\'y and S.~Samuel, {\it Phenomenological gravitational constraints on strings and higher-dimensional theories},  {\em Phys. Rev. Lett.} {\bf 63} (Jul, 1989) 224--227.

\bibitem{Agarwalla:2023wft}
S.~K. Agarwalla, S.~Das, S.~Sahoo, and P.~Swain, {\it {Constraining Lorentz invariance violation with next-generation long-baseline experiments}},  {\em JHEP} {\bf 07} (2023) 216, [\href{http://arxiv.org/abs/2302.12005}{{\tt arXiv:2302.12005}}].

\bibitem{KumarAgarwalla:2019gdj}
S.~Kumar~Agarwalla and M.~Masud, {\it {Can Lorentz invariance violation affect the sensitivity of deep underground neutrino experiment?}},  {\em Eur. Phys. J. C} {\bf 80} (2020), no.~8 716, [\href{http://arxiv.org/abs/1912.13306}{{\tt arXiv:1912.13306}}].

\bibitem{Kostelecky:2011gq}
A.~Kostelecky and M.~Mewes, {\it {Neutrinos with Lorentz-violating operators of arbitrary dimension}},  {\em Phys. Rev. D} {\bf 85} (2012) 096005, [\href{http://arxiv.org/abs/1112.6395}{{\tt arXiv:1112.6395}}].

\bibitem{PhysRevD.17.2369}
L.~Wolfenstein, {\it Neutrino oscillations in matter},  {\em Phys. Rev. D} {\bf 17} (May, 1978) 2369--2374.

\bibitem{Barenboim:2018lpo}
G.~Barenboim, C.~A. Ternes, and M.~T\'ortola, {\it {New physics vs new paradigms: distinguishing CPT violation from NSI}},  {\em Eur. Phys. J. C} {\bf 79} (2019), no.~5 390, [\href{http://arxiv.org/abs/1804.05842}{{\tt arXiv:1804.05842}}].

\bibitem{Diaz:2015dxa}
J.~S. Diaz, {\it {Correspondence between nonstandard interactions and CPT violation in neutrino oscillations}},  \href{http://arxiv.org/abs/1506.01936}{{\tt arXiv:1506.01936}}.

\bibitem{Barenboim:2018ctx}
G.~Barenboim, M.~Masud, C.~A. Ternes, and M.~T\'ortola, {\it {Exploring the intrinsic Lorentz-violating parameters at DUNE}},  {\em Phys. Lett. B} {\bf 788} (2019) 308--315, [\href{http://arxiv.org/abs/1805.11094}{{\tt arXiv:1805.11094}}].

\bibitem{Fiza:2022xfw}
N.~Fiza, N.~R. Khan~Chowdhury, and M.~Masud, {\it {Investigating Lorentz Invariance Violation with the long baseline experiment P2O}},  {\em JHEP} {\bf 01} (2023) 076, [\href{http://arxiv.org/abs/2206.14018}{{\tt arXiv:2206.14018}}].

\bibitem{MINOS:2008fnv}
{\bf MINOS} Collaboration, P.~Adamson et~al., {\it {Testing Lorentz Invariance and CPT Conservation with NuMI Neutrinos in the MINOS Near Detector}},  {\em Phys. Rev. Lett.} {\bf 101} (2008) 151601, [\href{http://arxiv.org/abs/0806.4945}{{\tt arXiv:0806.4945}}].

\bibitem{MINOS:2010kat}
{\bf MINOS} Collaboration, P.~Adamson et~al., {\it {A Search for Lorentz Invariance and CPT Violation with the MINOS Far Detector}},  {\em Phys. Rev. Lett.} {\bf 105} (2010) 151601, [\href{http://arxiv.org/abs/1007.2791}{{\tt arXiv:1007.2791}}].

\bibitem{MINOS:2012ozn}
{\bf MINOS} Collaboration, P.~Adamson et~al., {\it {Search for Lorentz invariance and CPT violation with muon antineutrinos in the MINOS Near Detector}},  {\em Phys. Rev. D} {\bf 85} (2012) 031101, [\href{http://arxiv.org/abs/1201.2631}{{\tt arXiv:1201.2631}}].

\bibitem{Dighe:2008bu}
A.~Dighe and S.~Ray, {\it {CPT violation in long baseline neutrino experiments: A Three flavor analysis}},  {\em Phys. Rev. D} {\bf 78} (2008) 036002, [\href{http://arxiv.org/abs/0802.0121}{{\tt arXiv:0802.0121}}].

\bibitem{Barenboim:2009ts}
G.~Barenboim and J.~D. Lykken, {\it {MINOS and CPT-violating neutrinos}},  {\em Phys. Rev. D} {\bf 80} (2009) 113008, [\href{http://arxiv.org/abs/0908.2993}{{\tt arXiv:0908.2993}}].

\bibitem{Rebel:2013vc}
B.~Rebel and S.~Mufson, {\it {The Search for Neutrino-Antineutrino Mixing Resulting from Lorentz Invariance Violation using Neutrino Interactions in MINOS}},  {\em Astropart. Phys.} {\bf 48} (2013) 78--81, [\href{http://arxiv.org/abs/1301.4684}{{\tt arXiv:1301.4684}}].

\bibitem{deGouvea:2017yvn}
A.~de~Gouv\^ea and K.~J. Kelly, {\it {Neutrino vs. Antineutrino Oscillation Parameters at DUNE and Hyper-Kamiokande}},  {\em Phys. Rev. D} {\bf 96} (2017), no.~9 095018, [\href{http://arxiv.org/abs/1709.06090}{{\tt arXiv:1709.06090}}].

\bibitem{Barenboim:2017ewj}
G.~Barenboim, C.~A. Ternes, and M.~T\'ortola, {\it {Neutrinos, DUNE and the world best bound on CPT invariance}},  {\em Phys. Lett. B} {\bf 780} (2018) 631--637, [\href{http://arxiv.org/abs/1712.01714}{{\tt arXiv:1712.01714}}].

\bibitem{Majhi:2019tfi}
R.~Majhi, S.~Chembra, and R.~Mohanta, {\it {Exploring the effect of Lorentz invariance violation with the currently running long-baseline experiments}},  {\em Eur. Phys. J. C} {\bf 80} (2020), no.~5 364, [\href{http://arxiv.org/abs/1907.09145}{{\tt arXiv:1907.09145}}].

\bibitem{Majhi:2022fed}
R.~Majhi, D.~K. Singha, M.~Ghosh, and R.~Mohanta, {\it {Distinguishing nonstandard interaction and Lorentz invariance violation at the Protvino to super-ORCA experiment}},  {\em Phys. Rev. D} {\bf 107} (2023), no.~7 075036, [\href{http://arxiv.org/abs/2212.07244}{{\tt arXiv:2212.07244}}].

\bibitem{T2K:2017ega}
{\bf T2K} Collaboration, K.~Abe et~al., {\it {Search for Lorentz and CPT violation using sidereal time dependence of neutrino flavor transitions over a short baseline}},  {\em Phys. Rev. D} {\bf 95} (2017), no.~11 111101, [\href{http://arxiv.org/abs/1703.01361}{{\tt arXiv:1703.01361}}].

\bibitem{MiniBooNE:2011pix}
{\bf MiniBooNE} Collaboration, A.~A. Aguilar-Arevalo et~al., {\it {Test of Lorentz and CPT violation with Short Baseline Neutrino Oscillation Excesses}},  {\em Phys. Lett. B} {\bf 718} (2013) 1303--1308, [\href{http://arxiv.org/abs/1109.3480}{{\tt arXiv:1109.3480}}].

\bibitem{Giunti:2010zs}
C.~Giunti and M.~Laveder, {\it {Hint of CPT Violation in Short-Baseline Electron Neutrino Disappearance}},  {\em Phys. Rev. D} {\bf 82} (2010) 113009, [\href{http://arxiv.org/abs/1008.4750}{{\tt arXiv:1008.4750}}].

\bibitem{DoubleChooz:2012eiq}
{\bf Double Chooz} Collaboration, Y.~Abe et~al., {\it {First Test of Lorentz Violation with a Reactor-based Antineutrino Experiment}},  {\em Phys. Rev. D} {\bf 86} (2012) 112009, [\href{http://arxiv.org/abs/1209.5810}{{\tt arXiv:1209.5810}}].

\bibitem{Diaz:2016fqd}
J.~S. Diaz and T.~Schwetz, {\it {Limits on CPT violation from solar neutrinos}},  {\em Phys. Rev. D} {\bf 93} (2016), no.~9 093004, [\href{http://arxiv.org/abs/1603.04468}{{\tt arXiv:1603.04468}}].

\bibitem{Hooper:2005jp}
D.~Hooper, D.~Morgan, and E.~Winstanley, {\it {Lorentz and CPT invariance violation in high-energy neutrinos}},  {\em Phys. Rev. D} {\bf 72} (2005) 065009, [\href{http://arxiv.org/abs/hep-ph/0506091}{{\tt hep-ph/0506091}}].

\bibitem{Tomar:2015fha}
G.~Tomar, S.~Mohanty, and S.~Pakvasa, {\it {Lorentz Invariance Violation and IceCube Neutrino Events}},  {\em JHEP} {\bf 11} (2015) 022, [\href{http://arxiv.org/abs/1507.03193}{{\tt arXiv:1507.03193}}].

\bibitem{Liao:2017yuy}
J.~Liao and D.~Marfatia, {\it {IceCube\textquoteright{}s astrophysical neutrino energy spectrum from CPT violation}},  {\em Phys. Rev. D} {\bf 97} (2018), no.~4 041302, [\href{http://arxiv.org/abs/1711.09266}{{\tt arXiv:1711.09266}}].

\bibitem{IceCube:2010fyu}
{\bf IceCube} Collaboration, R.~Abbasi et~al., {\it {Search for a Lorentz-violating sidereal signal with atmospheric neutrinos in IceCube}},  {\em Phys. Rev. D} {\bf 82} (2010) 112003, [\href{http://arxiv.org/abs/1010.4096}{{\tt arXiv:1010.4096}}].

\bibitem{Chatterjee:2014oda}
A.~Chatterjee, R.~Gandhi, and J.~Singh, {\it {Probing Lorentz and CPT Violation in a Magnetized Iron Detector using Atmospheric Neutrinos}},  {\em JHEP} {\bf 06} (2014) 045, [\href{http://arxiv.org/abs/1402.6265}{{\tt arXiv:1402.6265}}].

\bibitem{Sahoo:2021dit}
S.~Sahoo, A.~Kumar, and S.~K. Agarwalla, {\it {Probing Lorentz Invariance Violation with atmospheric neutrinos at INO-ICAL}},  {\em JHEP} {\bf 03} (2022) 050, [\href{http://arxiv.org/abs/2110.13207}{{\tt arXiv:2110.13207}}].

\bibitem{Datta:2003dg}
A.~Datta, R.~Gandhi, P.~Mehta, and S.~U. Sankar, {\it {Atmospheric neutrinos as a probe of CPT and Lorentz violation}},  {\em Phys. Lett. B} {\bf 597} (2004) 356--361, [\href{http://arxiv.org/abs/hep-ph/0312027}{{\tt hep-ph/0312027}}].

\bibitem{Super-Kamiokande:2014exs}
{\bf Super-Kamiokande} Collaboration, K.~Abe et~al., {\it {Test of Lorentz invariance with atmospheric neutrinos}},  {\em Phys. Rev. D} {\bf 91} (2015), no.~5 052003, [\href{http://arxiv.org/abs/1410.4267}{{\tt arXiv:1410.4267}}].

\bibitem{Arguelles:2019xgp}
C.~A. Arg\"uelles et~al., {\it {New opportunities at the next-generation neutrino experiments I: BSM neutrino physics and dark matter}},  {\em Rept. Prog. Phys.} {\bf 83} (2020), no.~12 124201, [\href{http://arxiv.org/abs/1907.08311}{{\tt arXiv:1907.08311}}].

\bibitem{Arguelles:2022tki}
C.~A. Arg\"uelles et~al., {\it {Snowmass white paper: beyond the standard model effects on neutrino flavor: Submitted to the proceedings of the US community study on the future of particle physics (Snowmass 2021)}},  {\em Eur. Phys. J. C} {\bf 83} (2023), no.~1 15, [\href{http://arxiv.org/abs/2203.10811}{{\tt arXiv:2203.10811}}].

\bibitem{nufit}
{\it Nufit 5.2},  {\em www.nu-fit.org} (2022).

\bibitem{Huber:2004ka}
P.~Huber, M.~Lindner, and W.~Winter, {\it {Simulation of long-baseline neutrino oscillation experiments with GLoBES (General Long Baseline Experiment Simulator)}},  {\em Comput. Phys. Commun.} {\bf 167} (2005) 195, [\href{http://arxiv.org/abs/hep-ph/0407333}{{\tt hep-ph/0407333}}].

\bibitem{Huber:2007ji}
P.~Huber, J.~Kopp, M.~Lindner, M.~Rolinec, and W.~Winter, {\it {New features in the simulation of neutrino oscillation experiments with GLoBES 3.0: General Long Baseline Experiment Simulator}},  {\em Comput. Phys. Commun.} {\bf 177} (2007) 432--438, [\href{http://arxiv.org/abs/hep-ph/0701187}{{\tt hep-ph/0701187}}].

\bibitem{Kopp:2007ne}
J.~Kopp, M.~Lindner, T.~Ota, and J.~Sato, {\it {Non-standard neutrino interactions in reactor and superbeam experiments}},  {\em Phys. Rev. D} {\bf 77} (2008) 013007, [\href{http://arxiv.org/abs/0708.0152}{{\tt arXiv:0708.0152}}].

\bibitem{Huber:2002mx}
P.~Huber, M.~Lindner, and W.~Winter, {\it {Superbeams versus neutrino factories}},  {\em Nucl. Phys. B} {\bf 645} (2002) 3--48, [\href{http://arxiv.org/abs/hep-ph/0204352}{{\tt hep-ph/0204352}}].

\bibitem{Fogli:2002pt}
G.~L. Fogli, E.~Lisi, A.~Marrone, D.~Montanino, and A.~Palazzo, {\it {Getting the most from the statistical analysis of solar neutrino oscillations}},  {\em Phys. Rev. D} {\bf 66} (2002) 053010, [\href{http://arxiv.org/abs/hep-ph/0206162}{{\tt hep-ph/0206162}}].

\bibitem{Agarwalla:2022xdo}
S.~K. Agarwalla, S.~Das, A.~Giarnetti, D.~Meloni, and M.~Singh, {\it {Enhancing sensitivity to leptonic CP violation using complementarity among DUNE, T2HK, and T2HKK}},  {\em Eur. Phys. J. C} {\bf 83} (2023), no.~8 694, [\href{http://arxiv.org/abs/2211.10620}{{\tt arXiv:2211.10620}}].

\bibitem{Denton:2020uda}
P.~B. Denton, J.~Gehrlein, and R.~Pestes, {\it {$CP$ -Violating Neutrino Nonstandard Interactions in Long-Baseline-Accelerator Data}},  {\em Phys. Rev. Lett.} {\bf 126} (2021), no.~5 051801, [\href{http://arxiv.org/abs/2008.01110}{{\tt arXiv:2008.01110}}].

\bibitem{deGouvea:2015ndi}
A.~de~Gouvêa and K.~J. Kelly, {\it {Non-standard Neutrino Interactions at DUNE}},  {\em Nucl. Phys. B} {\bf 908} (2016) 318--335, [\href{http://arxiv.org/abs/1511.05562}{{\tt arXiv:1511.05562}}].

\bibitem{Dev:2019anc}
{\em {Neutrino Non-Standard Interactions: A Status Report}}, vol.~2, 2019.

\bibitem{Bakhti:2020fde}
P.~Bakhti and M.~Rajaee, {\it {Sensitivities of future reactor and long-baseline neutrino experiments to NSI}},  {\em Phys. Rev. D} {\bf 103} (2021), no.~7 075003, [\href{http://arxiv.org/abs/2010.12849}{{\tt arXiv:2010.12849}}].

\bibitem{Giarnetti:2021wur}
A.~Giarnetti and D.~Meloni, {\it {New Sources of Leptonic CP Violation at the DUNE Neutrino Experiment}},  {\em Universe} {\bf 7} (2021), no.~7 240, [\href{http://arxiv.org/abs/2106.00030}{{\tt arXiv:2106.00030}}].

\bibitem{Ge:2018uhz}
S.-F. Ge and S.~J. Parke, {\it {Scalar Nonstandard Interactions in Neutrino Oscillation}},  {\em Phys. Rev. Lett.} {\bf 122} (2019), no.~21 211801, [\href{http://arxiv.org/abs/1812.08376}{{\tt arXiv:1812.08376}}].

\bibitem{Denton:2022pxt}
P.~B. Denton, A.~Giarnetti, and D.~Meloni, {\it {How to identify different new neutrino oscillation physics scenarios at DUNE}},  {\em JHEP} {\bf 02} (2023) 210, [\href{http://arxiv.org/abs/2210.00109}{{\tt arXiv:2210.00109}}].

\bibitem{ESSnuSB:2023lbg}
{\bf ESSnuSB} Collaboration, J.~Aguilar et~al., {\it {Study of non-standard interaction mediated by a scalar field at ESSnuSB experiment}},  \href{http://arxiv.org/abs/2310.10749}{{\tt arXiv:2310.10749}}.

\bibitem{Gupta:2023wct}
A.~Gupta, D.~Majumdar, and S.~Prakash, {\it {Neutrino oscillation measurements with JUNO in the presence of scalar NSI}},  \href{http://arxiv.org/abs/2306.07343}{{\tt arXiv:2306.07343}}.

\bibitem{Sarker:2023qzp}
A.~Sarker, A.~Medhi, D.~Bezboruah, M.~M. Devi, and D.~Dutta, {\it {Impact of scalar NSI on the neutrino mass hierarchy sensitivity at DUNE, T2HK and T2HKK}},  \href{http://arxiv.org/abs/2309.12249}{{\tt arXiv:2309.12249}}.

\bibitem{Grifols:1993rs}
J.~A. Grifols, E.~Masso, and S.~Peris, {\it {Supernova neutrinos as probes of long range nongravitational interactions of dark matter}},  {\em Astropart. Phys.} {\bf 2} (1994) 161--165.

\bibitem{Grifols:1996fk}
J.~A. Grifols, E.~Masso, and R.~Toldra, {\it {Majorana neutrinos and long range forces}},  {\em Phys. Lett. B} {\bf 389} (1996) 563--565, [\href{http://arxiv.org/abs/hep-ph/9606377}{{\tt hep-ph/9606377}}].

\bibitem{Grifols:2003gy}
J.~A. Grifols and E.~Masso, {\it {Neutrino oscillations in the sun probe long range leptonic forces}},  {\em Phys. Lett. B} {\bf 579} (2004) 123--126, [\href{http://arxiv.org/abs/hep-ph/0311141}{{\tt hep-ph/0311141}}].

\bibitem{Mishra:2024riq}
P.~Mishra, R.~Majhi, S.~K. Pusty, M.~Ghosh, and R.~Mohanta, {\it {Study of Long Range Force in P2SO and T2HKK}},  \href{http://arxiv.org/abs/2402.19178}{{\tt arXiv:2402.19178}}.

\bibitem{Agarwalla:2024ylc}
S.~K. Agarwalla, M.~Bustamante, M.~Singh, and P.~Swain, {\it {A plethora of long-range neutrino interactions probed by DUNE and T2HK}},  \href{http://arxiv.org/abs/2404.02775}{{\tt arXiv:2404.02775}}.

\bibitem{Agarwalla:2023sng}
S.~K. Agarwalla, M.~Bustamante, S.~Das, and A.~Narang, {\it {Present and future constraints on flavor-dependent long-range interactions of high-energy astrophysical neutrinos}},  {\em JHEP} {\bf 08} (2023) 113, [\href{http://arxiv.org/abs/2305.03675}{{\tt arXiv:2305.03675}}].

\bibitem{Chatterjee:2015gta}
S.~S. Chatterjee, A.~Dasgupta, and S.~K. Agarwalla, {\it {Exploring Flavor-Dependent Long-Range Forces in Long-Baseline Neutrino Oscillation Experiments}},  {\em JHEP} {\bf 12} (2015) 167, [\href{http://arxiv.org/abs/1509.03517}{{\tt arXiv:1509.03517}}].

\bibitem{Coloma:2020gfv}
P.~Coloma, M.~C. Gonzalez-Garcia, and M.~Maltoni, {\it {Neutrino oscillation constraints on U(1)' models: from non-standard interactions to long-range forces}},  {\em JHEP} {\bf 01} (2021) 114, [\href{http://arxiv.org/abs/2009.14220}{{\tt arXiv:2009.14220}}]. [Erratum: JHEP 11, 115 (2022)].

\bibitem{PhysRevD.43.R22}
X.~G. He, G.~C. Joshi, H.~Lew, and R.~R. Volkas, {\it New-${Z}^{\ensuremath{'}}$ phenomenology},  {\em Phys. Rev. D} {\bf 43} (Jan, 1991) R22--R24.

\bibitem{PhysRevD.44.2118}
X.-G. He, G.~C. Joshi, H.~Lew, and R.~R. Volkas, {\it Simplest ${Z}^{\ensuremath{'}}$ model},  {\em Phys. Rev. D} {\bf 44} (Oct, 1991) 2118--2132.

\bibitem{Foot:1994vd}
R.~Foot, X.~G. He, H.~Lew, and R.~R. Volkas, {\it {Model for a light Z-prime boson}},  {\em Phys. Rev. D} {\bf 50} (1994) 4571--4580, [\href{http://arxiv.org/abs/hep-ph/9401250}{{\tt hep-ph/9401250}}].

\bibitem{Asai:2018ocx}
K.~Asai, K.~Hamaguchi, N.~Nagata, S.-Y. Tseng, and K.~Tsumura, {\it {Minimal Gauged U(1)$_{L_\alpha - L_\beta}$ Models Driven into a Corner}},  {\em Phys. Rev. D} {\bf 99} (2019), no.~5 055029, [\href{http://arxiv.org/abs/1811.07571}{{\tt arXiv:1811.07571}}].

\bibitem{Asai:2017ryy}
K.~Asai, K.~Hamaguchi, and N.~Nagata, {\it {Predictions for the neutrino parameters in the minimal gauged U(1)$_{L_\mu-L_\tau}$ model}},  {\em Eur. Phys. J. C} {\bf 77} (2017), no.~11 763, [\href{http://arxiv.org/abs/1705.00419}{{\tt arXiv:1705.00419}}].

\bibitem{Lou:2024fvw}
Y.~Lou and T.~Nomura, {\it {Neutrino observables in gauged $U(1)_{L_\alpha-L_\beta}$ models with two Higgs doublet and one singlet scalars}},  \href{http://arxiv.org/abs/2406.01030}{{\tt arXiv:2406.01030}}.

\bibitem{Bustamante:2018mzu}
M.~Bustamante and S.~K. Agarwalla, {\it {Universe's Worth of Electrons to Probe Long-Range Interactions of High-Energy Astrophysical Neutrinos}},  {\em Phys. Rev. Lett.} {\bf 122} (2019), no.~6 061103, [\href{http://arxiv.org/abs/1808.02042}{{\tt arXiv:1808.02042}}].

\bibitem{Wise:2018rnb}
M.~B. Wise and Y.~Zhang, {\it {Lepton Flavorful Fifth Force and Depth-dependent Neutrino Matter Interactions}},  {\em JHEP} {\bf 06} (2018) 053, [\href{http://arxiv.org/abs/1803.00591}{{\tt arXiv:1803.00591}}].

\bibitem{Singh:2023nek}
M.~Singh, M.~Bustamante, and S.~K. Agarwalla, {\it {Flavor-dependent long-range neutrino interactions in DUNE \& T2HK: alone they constrain, together they discover}},  {\em JHEP} {\bf 08} (2023) 101, [\href{http://arxiv.org/abs/2305.05184}{{\tt arXiv:2305.05184}}].

\bibitem{Khatun:2018lzs}
A.~Khatun, T.~Thakore, and S.~Kumar~Agarwalla, {\it {Can INO be Sensitive to Flavor-Dependent Long-Range Forces?}},  {\em JHEP} {\bf 04} (2018) 023, [\href{http://arxiv.org/abs/1801.00949}{{\tt arXiv:1801.00949}}].

\bibitem{Agarwalla:2021zfr}
S.~K. Agarwalla, S.~Das, M.~Masud, and P.~Swain, {\it {Evolution of neutrino mass-mixing parameters in matter with non-standard interactions}},  {\em JHEP} {\bf 11} (2021) 094, [\href{http://arxiv.org/abs/2103.13431}{{\tt arXiv:2103.13431}}].

\bibitem{Joshipura:2003jh}
A.~S. Joshipura and S.~Mohanty, {\it {Constraints on flavor dependent long range forces from atmospheric neutrino observations at super-Kamiokande}},  {\em Phys. Lett. B} {\bf 584} (2004) 103--108, [\href{http://arxiv.org/abs/hep-ph/0310210}{{\tt hep-ph/0310210}}].

\bibitem{Bandyopadhyay:2006uh}
A.~Bandyopadhyay, A.~Dighe, and A.~S. Joshipura, {\it {Constraints on flavor-dependent long range forces from solar neutrinos and KamLAND}},  {\em Phys. Rev. D} {\bf 75} (2007) 093005, [\href{http://arxiv.org/abs/hep-ph/0610263}{{\tt hep-ph/0610263}}].

\bibitem{Gonzalez-Garcia:2013usa}
M.~C. Gonzalez-Garcia and M.~Maltoni, {\it {Determination of matter potential from global analysis of neutrino oscillation data}},  {\em JHEP} {\bf 09} (2013) 152, [\href{http://arxiv.org/abs/1307.3092}{{\tt arXiv:1307.3092}}].

\bibitem{Honda:2007wv}
M.~Honda, Y.~Kao, N.~Okamura, A.~Pronin, and T.~Takeuchi, {\it {Constraints on New Physics from Long Baseline Neutrino Oscillation Experiments}},  \href{http://arxiv.org/abs/0707.4545}{{\tt arXiv:0707.4545}}.

\bibitem{Farzan:2018pnk}
Y.~Farzan and S.~Palomares-Ruiz, {\it {Flavor of cosmic neutrinos preserved by ultralight dark matter}},  {\em Phys. Rev. D} {\bf 99} (2019), no.~5 051702, [\href{http://arxiv.org/abs/1810.00892}{{\tt arXiv:1810.00892}}].

\bibitem{IceCube-Gen2:2020qha}
{\bf IceCube-Gen2} Collaboration, M.~G. Aartsen et~al., {\it {IceCube-Gen2: the window to the extreme Universe}},  {\em J. Phys. G} {\bf 48} (2021), no.~6 060501, [\href{http://arxiv.org/abs/2008.04323}{{\tt arXiv:2008.04323}}].

\bibitem{McMillan_2011}
P.~J. McMillan, {\it Mass models of the milky way},  {\em Monthly Notices of the Royal Astronomical Society} {\bf 414} (2011), no.~3 2446--2457.

\bibitem{Miller_2013}
M.~J. Miller and J.~N. Bregman, {\it {The Structure of the Milky Way's Hot Gas Halo}},  {\em Astrophys. J.} {\bf 770} (2013) 118, [\href{http://arxiv.org/abs/1305.2430}{{\tt arXiv:1305.2430}}].

\bibitem{Baryakhtar_2017}
M.~Baryakhtar, R.~Lasenby, and M.~Teo, {\it {Black Hole Superradiance Signatures of Ultralight Vectors}},  {\em Phys. Rev. D} {\bf 96} (2017), no.~3 035019, [\href{http://arxiv.org/abs/1704.05081}{{\tt arXiv:1704.05081}}].

\bibitem{Arkani_Hamed_2007}
N.~Arkani-Hamed, L.~Motl, A.~Nicolis, and C.~Vafa, {\it {The String landscape, black holes and gravity as the weakest force}},  {\em JHEP} {\bf 06} (2007) 060, [\href{http://arxiv.org/abs/hep-th/0601001}{{\tt hep-th/0601001}}].

\bibitem{deGouvea:2016qpx}
A.~de~Gouv\^ea, {\it {Neutrino Mass Models}},  {\em Ann. Rev. Nucl. Part. Sci.} {\bf 66} (2016) 197--217.

\bibitem{King_2003}
S.~F. King, {\it {Neutrino mass models}},  {\em Rept. Prog. Phys.} {\bf 67} (2004) 107--158, [\href{http://arxiv.org/abs/hep-ph/0310204}{{\tt hep-ph/0310204}}].

\bibitem{Arkani-Hamed:1998wuz}
N.~Arkani-Hamed, S.~Dimopoulos, G.~R. Dvali, and J.~March-Russell, {\it {Neutrino masses from large extra dimensions}},  {\em Phys. Rev. D} {\bf 65} (2001) 024032, [\href{http://arxiv.org/abs/hep-ph/9811448}{{\tt hep-ph/9811448}}].

\bibitem{Dienes_1999}
K.~R. Dienes, E.~Dudas, and T.~Gherghetta, {\it {Neutrino oscillations without neutrino masses or heavy mass scales: A Higher dimensional seesaw mechanism}},  {\em Nucl. Phys. B} {\bf 557} (1999) 25, [\href{http://arxiv.org/abs/hep-ph/9811428}{{\tt hep-ph/9811428}}].

\bibitem{Dvali_1999}
G.~R. Dvali and A.~Y. Smirnov, {\it {Probing large extra dimensions with neutrinos}},  {\em Nucl. Phys. B} {\bf 563} (1999) 63--81, [\href{http://arxiv.org/abs/hep-ph/9904211}{{\tt hep-ph/9904211}}].

\bibitem{Mohapatra_2001}
R.~N. Mohapatra and A.~Perez-Lorenzana, {\it {Three flavor neutrino oscillations in models with large extra dimensions}},  {\em Nucl. Phys. B} {\bf 593} (2001) 451--470, [\href{http://arxiv.org/abs/hep-ph/0006278}{{\tt hep-ph/0006278}}].

\bibitem{Barbieri_2000}
R.~Barbieri, P.~Creminelli, and A.~Strumia, {\it {Neutrino oscillations from large extra dimensions}},  {\em Nucl. Phys. B} {\bf 585} (2000) 28--44, [\href{http://arxiv.org/abs/hep-ph/0002199}{{\tt hep-ph/0002199}}].

\bibitem{Davoudiasl_2002}
H.~Davoudiasl, P.~Langacker, and M.~Perelstein, {\it {Constraints on large extra dimensions from neutrino oscillation experiments}},  {\em Phys. Rev. D} {\bf 65} (2002) 105015, [\href{http://arxiv.org/abs/hep-ph/0201128}{{\tt hep-ph/0201128}}].

\bibitem{Arkani_Hamed_1998_1}
N.~Arkani-Hamed, S.~Dimopoulos, and G.~R. Dvali, {\it {The Hierarchy problem and new dimensions at a millimeter}},  {\em Phys. Lett. B} {\bf 429} (1998) 263--272, [\href{http://arxiv.org/abs/hep-ph/9803315}{{\tt hep-ph/9803315}}].

\bibitem{Arkani_Hamed_1999}
N.~Arkani-Hamed, S.~Dimopoulos, and G.~R. Dvali, {\it {Phenomenology, astrophysics and cosmology of theories with submillimeter dimensions and TeV scale quantum gravity}},  {\em Phys. Rev. D} {\bf 59} (1999) 086004, [\href{http://arxiv.org/abs/hep-ph/9807344}{{\tt hep-ph/9807344}}].

\bibitem{Esmaili_2014}
A.~Esmaili, O.~L.~G. Peres, and Z.~Tabrizi, {\it {Probing Large Extra Dimensions With IceCube}},  {\em JCAP} {\bf 12} (2014) 002, [\href{http://arxiv.org/abs/1409.3502}{{\tt arXiv:1409.3502}}].

\bibitem{Machado_2011}
P.~A.~N. Machado, H.~Nunokawa, and R.~Zukanovich~Funchal, {\it {Testing for Large Extra Dimensions with Neutrino Oscillations}},  {\em Phys. Rev. D} {\bf 84} (2011) 013003, [\href{http://arxiv.org/abs/1101.0003}{{\tt arXiv:1101.0003}}].

\bibitem{Machado_2012}
P.~A.~N. Machado, H.~Nunokawa, F.~A.~P. dos Santos, and R.~Z. Funchal, {\it {Bulk Neutrinos as an Alternative Cause of the Gallium and Reactor Anti-neutrino Anomalies}},  {\em Phys. Rev. D} {\bf 85} (2012) 073012, [\href{http://arxiv.org/abs/1107.2400}{{\tt arXiv:1107.2400}}].

\bibitem{Basto_Gonzalez_2013}
V.~S. Basto-Gonzalez, A.~Esmaili, and O.~L.~G. Peres, {\it {Kinematical Test of Large Extra Dimension in Beta Decay Experiments}},  {\em Phys. Lett. B} {\bf 718} (2013) 1020--1023, [\href{http://arxiv.org/abs/1205.6212}{{\tt arXiv:1205.6212}}].

\bibitem{Girardi_2014}
I.~Girardi and D.~Meloni, {\it {Constraining new physics scenarios in neutrino oscillations from Daya Bay data}},  {\em Phys. Rev. D} {\bf 90} (2014), no.~7 073011, [\href{http://arxiv.org/abs/1403.5507}{{\tt arXiv:1403.5507}}].

\bibitem{Rodejohann_2014}
W.~Rodejohann and H.~Zhang, {\it {Signatures of Extra Dimensional Sterile Neutrinos}},  {\em Phys. Lett. B} {\bf 737} (2014) 81--89, [\href{http://arxiv.org/abs/1407.2739}{{\tt arXiv:1407.2739}}].

\bibitem{Carena:2017qhd}
M.~Carena, Y.-Y. Li, C.~S. Machado, P.~A.~N. Machado, and C.~E.~M. Wagner, {\it {Neutrinos in Large Extra Dimensions and Short-Baseline $\nu_e$ Appearance}},  {\em Phys. Rev. D} {\bf 96} (2017), no.~9 095014, [\href{http://arxiv.org/abs/1708.09548}{{\tt arXiv:1708.09548}}].

\bibitem{Stenico_2018}
G.~V. Stenico, D.~V. Forero, and O.~L.~G. Peres, {\it {A Short Travel for Neutrinos in Large Extra Dimensions}},  {\em JHEP} {\bf 11} (2018) 155, [\href{http://arxiv.org/abs/1808.05450}{{\tt arXiv:1808.05450}}].

\bibitem{Arg_elles_2020}
C.~A. Arg\"uelles et~al., {\it {New opportunities at the next-generation neutrino experiments I: BSM neutrino physics and dark matter}},  {\em Rept. Prog. Phys.} {\bf 83} (2020), no.~12 124201, [\href{http://arxiv.org/abs/1907.08311}{{\tt arXiv:1907.08311}}].

\bibitem{bastogonzalez2022shortbaseline}
V.~S. Basto-Gonzalez, D.~V. Forero, C.~Giunti, A.~A. Quiroga, and C.~A. Ternes, {\it {Short-baseline oscillation scenarios at JUNO and TAO}},  2022.

\bibitem{Forero_2022}
D.~V. Forero, C.~Giunti, C.~A. Ternes, and O.~Tyagi, {\it {Large extra dimensions and neutrino experiments}},  {\em Phys. Rev. D} {\bf 106} (2022), no.~3 035027, [\href{http://arxiv.org/abs/2207.02790}{{\tt arXiv:2207.02790}}].

\bibitem{Berryman_2016}
J.~M. Berryman, A.~de~Gouv\^ea, K.~J. Kelly, O.~L.~G. Peres, and Z.~Tabrizi, {\it {Large, Extra Dimensions at the Deep Underground Neutrino Experiment}},  {\em Phys. Rev. D} {\bf 94} (2016), no.~3 033006, [\href{http://arxiv.org/abs/1603.00018}{{\tt arXiv:1603.00018}}].

\end{thebibliography}\endgroup
\end{document}